\documentclass[12pt]{iopart}
\pdfoutput=1
\usepackage[utf8]{inputenc}
\usepackage[english]{babel}
\usepackage[T1]{fontenc}

\usepackage{iopams}

\expandafter\let\csname equation*\endcsname\relax

\expandafter\let\csname endequation*\endcsname\relax

\usepackage{amsmath}

\usepackage{lineno}

\usepackage[sort&compress,numbers]{natbib}
\usepackage{hyperref}

\usepackage{tikz}
\usepackage{lipsum}
\usepackage{bbm}

\usepackage{verbatim}
\usepackage{dsfont}
\usepackage{amssymb}
\usepackage{graphicx}
\usepackage{esint}
\usepackage{appendix}
\usepackage{xcolor}
\usepackage{braket}
\usepackage[mathscr]{eucal}
\usepackage{color}
\graphicspath{{figures/}}

\newcommand{\GJC}{\hat{G}_{JC}}

\begin{document}

\title[From the Jaynes-Cummings model to non-Abelian gauge theories]{From the Jaynes-Cummings model to non-Abelian gauge theories: a guided tour for the quantum engineer}

\author{Valentin Kasper}
\address{ICFO - Institut de Ciencies Fotoniques, The Barcelona Institute of Science and Technology, Av. Carl Friedrich Gauss 3, 08860 Castelldefels (Barcelona), Spain}
\address{Department of Physics, Harvard University, Cambridge, MA, 02138, US}
\author{Gediminas Juzeli\={u}nas}
\address{Institute of Theoretical Physics and Astronomy, Vilnius University, Saul\.etekio 3, 10257 Vilnius, Lithuania }
\author{Maciej Lewenstein}
\address{ICFO - Institut de Ciencies Fotoniques, The Barcelona Institute of Science and Technology, Av. Carl Friedrich Gauss 3, 08860 Castelldefels (Barcelona), Spain}
\address{ICREA, Pg. Lluis Companys 23, 08010 Barcelona, Spain}
\author{Fred Jendrzejewski }
\address{Kirchhoff-Institut  f\"ur  Physik,  Im  Neuenheimer  Feld  227,  69120  Heidelberg, Germany}
\author{Erez Zohar}
\address{Racah Institute of Physics, The Hebrew University of Jerusalem, Givat Ram, Jerusalem 91904, Israel}

\maketitle

\begin{abstract}
The design of quantum many body systems, which have to fulfill an extensive number of constraints, appears as a formidable challenge within the field of quantum simulation. Lattice gauge theories are a particular important class of quantum systems with an extensive number of local constraints and play a central role in high energy physics, condensed matter and quantum information. Whereas recent experimental progress points towards the feasibility of large-scale quantum simulation of Abelian gauge theories, the quantum simulation of non-Abelian gauge theories appears still elusive. In this paper we present minimal non-Abelian lattice gauge theories, whereby we introduce the necessary formalism in well-known Abelian gauge theories, such as the Jaynes-Cumming model.  In particular, we show that certain minimal non-Abelian lattice gauge theories can be mapped to three or four level systems, for which the design of a quantum simulator is standard with current technologies. Further we give an upper bound for the Hilbert space dimension of a one dimensional SU(2) lattice gauge theory, and argue that the implementation with current digital quantum computer appears feasible. 
\end{abstract}

\section{Introduction and Motivation}\label{Sec:Motivation}

Gauge theories are of fundamental importance to understand vastly different states of matter ranging from particle physics \cite{Yang1954} through strongly correlated materials \cite{Levin2005} to quantum information processing \cite{Kitaev2006}. In condensed matter physics, gauge theories appear as effective descriptions \cite{Wen_book}, whereas in particle physics gauge invariance is usually postulated \cite{Weinberg_QFT}.  One of the most important gauge theories is quantum electrodynamics (QED), which describes the interaction of photons with charged particles and is a representative of an Abelian gauge theory. In contrast, quantum chromodynamics, which models the strong interactions between quarks and gluons, is a non-Abelian gauge theory. In order to predict the phases of quarks and gluons one has to enter the non-perturbative regime of quantum chromodynamics. For this purpose, lattice gauge theories provide a framework to numerically study gauge theories on a discretized space \cite{Kogut1975, Kogut1979} or space-time \cite{Wilson1974}. Monte Carlo techniques are particularly successful in predicting the hadron spectrum \cite{Aoki2017}. However, despite the significant knowledge gained from Monte Carlo simulations, fundamental challenges remain. For example, frequently the Monte Carlo approach is not able to study systems with a finite fermionic density due to the well-known sign problem \cite{Troyer2005}. Similarly, Monte Carlo simulations are not able to describe the unitary real-time evolution. However, in order to access real-time evolution or strongly coupled quantum many body systems, quantum simulators appear as a promising alternative to numerical calculations. Quantum simulators \cite{Feynman1982} are special purpose devices which realize a specific quantum system. A quantum simulator  for quantum chromodynamics would have the great prospect to address challenging problems such as the physics of confinement and deconfinement, chiral symmetry breaking at finite fermion density, color-superconductivity, or the collisions of heavy-ions.  

These prospects of quantum simulating lattice gauge theories led to considerable attention in recent years \cite{Wiese2013, Zohar2016, Dalmonte2016, Banuls_review_2019}. The first proposals for cold atomic quantum simulators of lattice gauge theories were focused on Abelian \cite{Zohar2011, Banerjee2012, Tagliacozzo_abelian_2013, Zohar2012, Zohar_PRLa_2013} and later non-Abelian \cite{Tagliacozzo_nonabelian_2013, Zohar2013, Zohar_PRL_2013, Banerjee2013} local gauge invariance. These proposals contained various approaches to implement gauge invariant interactions among ultracold atoms in optical lattices. In the following years, several groups joined the research effort. This effort resulted in  new approaches and implementation schemes for analog and digital quantum simulation of lattice gauge theories covering various gauge groups, using different implementation techniques, and platforms such as atoms with spin-changing collisions, alkaline-earth atoms, Rydberg atoms in optical tweezers, dissipative quantum systems, superconducting qubits, ion experiment to name a few \cite{Stannigel2013, Zohar_digital_PRL_2017, Hauke2014, Marcos2013, Marcos2014, Zohar_digital_PRA_2017, Bender2018, Laflamme2016, Dutta2017, Gonzalez-Cuadra2017, Rico2018, Surace2019, Celi2019, Kasper2017, Zache2018, Klco2018, Kaplan2018, Stryker2019, Klco2019, Mezzacapo2015, Bazavov2015, Zhang2018, Magnifico2019, Kuno2015, Notarnicola2015, Zohar2020}. Besides the quest for more efficient and 
practicable approaches to build quantum simulators of gauge theories, the reliability of coming quantum simulators of gauge theories is a pressing theoretical and experimental question \cite{Halimeh2019,Halimeh2020}.

First experimental results on Abelian gauge theories were obtained with trapped ion systems \cite{Martinez2016, Kokail2019}. In addition, experiments with ultracold atoms realize the building blocks of discrete \cite{Schweizer2019, Gorg2019} and continuous Abelian symmetries \cite{Mil2020, Yang2020}. However, the experimental progress towards the quantum simulation of non-Abelian lattice gauge theories, such as SU(2), is yet missing. Therefore, it is desirable to identify the minimal, yet non-trivial, non-Abelian lattice gauge models which can be  implemented in current experiments.

In this work we offer a guide to quantum engineers interested in lattice gauge theories, by discussing lattice gauge theories in familiar terms of quantum optics. In particular, we consider minimal instances of Abelian and non-Abelian lattice gauge theories, and analyze them in an analogy to the Jaynes-Cummings model. We show that a minimal instance of a non-Abelian SU(2) lattice gauge theory can be mapped to a two-qubit system, which in turn can be quantum simulated or simulated on quantum computation architectures. We extend the discussion to arbitrarily long one dimensional systems and argue that quantum simulation of minimal instances of non-Abelian lattice gauge theories is possible with current technologies.

The paper is structured as follows. In section \ref{sec:2} we start with considering a minimal U(1) lattice gauge theory, which is frequently used in quantum optics and solid state physics: the Jaynes-Cumming model. In section \ref{sec:3} we show that compact quantum electrodynamics appears as a simple modification of the Jaynes-Cumming model and in section  \ref{sec:4} we generalize our minimal Abelian lattice gauge  to a minimal non-Abelian lattice gauge theory. In particular, we show that the minimal non-Abelian lattice gauge theories can be quantum simulated with a three or four level system respectively. Finally we discuss the dimension of the constrained Hilbert space of one dimensional U(1) and SU(2) lattice gauge theories and discuss the importance of the basis vectors, which are used to quantum simulate non-Abelian gauge theories.

\section{Cavity QED within the Jaynes-Cummings model \label{sec:2}}
The Jaynes-Cummings model (JCM) describes the interaction of
a two-level atom with a quantized photon-mode hosted by an optical cavity. 
It is used to investigate the interaction of an atom with the quantized electromagnetic field in order to understand e.g. spontaneous emission or absorption of photons in a cavity \cite{Raimond2001}. 
The JCM is not only used in atomic physics \cite{Meekhof1996,Wineland2013}, but also solid-state physics \cite{Wallraff2004}, and has wide ranging experimental and theoretical relevance e.g. it is encountered as the simplest model of supersymmetric quantum mechanics \cite{Andreev1989}. 

\begin{figure}\centering{\includegraphics[width=\columnwidth]{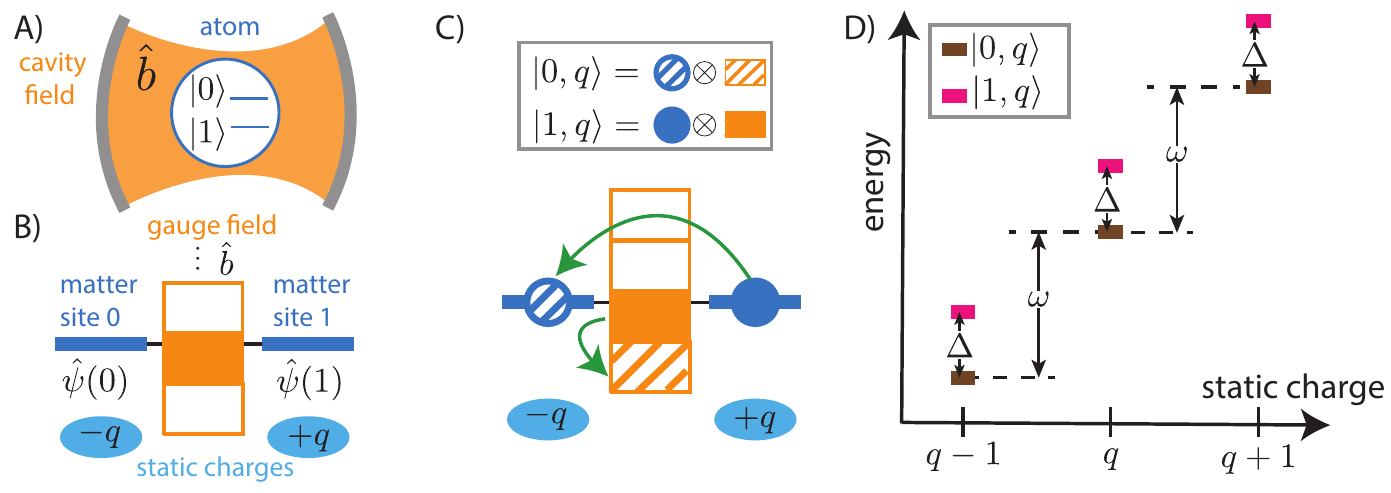}
\caption{\label{Fig:JaynesCummings} The Jaynes-Cummings model (JCM). A) In its original version the JCM describes a cavity mode $\hat{b}$ (yellow), which is coupled to a two-level system (blue) characterized by the ground  and excited states, $\ket{1}$ and $\ket{0}$ respectively. B) The JCM can be reinterpreted in terms of two fermionic sites, which are coupled in a gauge-invariant way by the bosonic link. Each doublet can be labeled by a static charge $q$, which is positioned on the two matter sites. C) The two states spanning the doublet of the JCM. D) The gauge invariant coupling leads to the formation of doublets in the Jaynes-Cummings ladder.}
}
\end{figure}

The Hilbert space of the JCM is the tensor product of the Hilbert space of one photon mode and the Hilbert space of one two level atom. The basis of the photon Hilbert
space is given by eigenstates of the occupation number of the photonic mode $\ket{n}_B$, where 
$n$ is a non-negative integer. The basis for the two-level atom is given by
the ground state  $\ket{1}$ and the excited state  $\ket{0}$, as illustrated in Fig.~\ref{Fig:JaynesCummings}A. The  \emph{lowering operator}  $\hat{b}$  describes annihilation of a photon, i.e.   $\hat{b}\ket{n}_B = \sqrt{n}\ket{n-1}_B$, while the Hermitian conjugate  \emph{raising operator} $\hat{b}^{\dagger}$ creates a photon: $\hat{b}^{\dagger}\ket{n}_B = \sqrt{n+1}\ket{n+1}_B$.  The Hamiltonian of the JCM is given by
\begin{align}
\hat{H}_{JC}= \omega_c \hat{b}^{\dagger} \hat{b}+  \frac{\omega_a}{2} (\ket{0}\!\bra{0} - \ket{1}\!\bra{1}) +\frac{\Omega_0}{2}\left(  \hat{b}^{\dagger} \ket{1}\!\bra{0} + h.c. \right) \,.
\end{align}
The first term of the Hamiltonian is the energy of the photons in the cavity, the second is the energy-gain of an electron transitioning from the ground state to 
the excited state, and $\Omega_0$ is the Rabi-frequency, which quantifies the
coupling between the photon and the electron of the two-level atom. The third term describes the excitation of the atom through the absorption of a photon and vice versa. From this coupling term one can deduce, that the state $\ket{0, n}$ couples exclusively to the state $\ket{1, n+1}$ and we can decompose each state into such  doublets. Decomposing the Hilbert space into doublets given by $\{\ket{0, n}, \ket{1, n+1}\}$ brings the Hamiltonian into a block diagonal form. For each $n\geq1$, we obtain a pair of eigenstates and eigenenergies, which form \emph{the Jaynes-Cummings ladder}. Each block of the Hamiltonian entails important phenomena such as quantum Rabi oscillations \cite{Brune1996a}. The block diagonal structure of the Hamiltonian originates from a local (gauge) U(1)		 symmetry, which we will discuss in the following.

\subsection{JCM using the fermionic Schwinger representation}

In this section we investigate the JCM focusing on its global and local symmetries. The discussion will allow us to formulate and study extended systems with Abelian and non-Abelian local symmetries. First we define the Hilbert space. The two level atom can be represented by one fermionic particle in two fermionic modes (fermionic Schwinger representation). 
The ground state $\ket{1}$ is identified with $\hat{\psi}^{\dagger}(1)\ket{\mathrm{vac}}_F$ 
and the excited state $\ket{0}$ with $\hat{\psi}^{\dagger}(0)\ket{\mathrm{vac}}_F$, where $\hat{\psi}^{\dagger}(0)$ and $\hat{\psi}^{\dagger}(1)$ are the creation operators of the fermionic modes and $\ket{\mathrm{vac}}_F$ is the fermionic Fock vacuum.  The bosonic and fermionic nature of the operators $\hat{b}$ and $\hat{\psi}(x)$ is manifested in the commutation and anti-commutation relations
\begin{equation}
[\hat{b},\hat{b}^{\dagger}]=1\,,\qquad\{\hat{\psi}(x),\hat{\psi}^{\dagger}(y)\}=\delta(x,y)\,,\label{eq:commutation-rel-JCM}
\end{equation}
with $x,y \in \{0,1\}$, and $\left[\hat{A}, \hat{B}\right]= \hat{A} \hat{B}- \hat{B} \hat{A}$ denotes a commutator, whereas $\left\{\hat{A},\hat{B}\right\}=\hat{A}\hat{B}+ \hat{B} \hat{A}$ is the anti-commutator. The label $x$ of the fermionic operator $\hat{\psi}(x)$ denotes the ground or
excited state of the two state atom, but $x$ can also be interpreted as
the position of a fermionic atom on a lattice site.  Using the fermionic Schwinger
representation the JCM can be written as
\begin{align}
\hat{H}_{JC}= \omega_c \hat{b}^{\dagger} \hat{b}+  \frac{\omega_a}{2} \left[\hat{\psi}^\dag(0) \hat{\psi}(0) - \hat{\psi}^\dag(1) \hat{\psi}(1)\right]+\frac{ \Omega_0}{2}\left[ \hat{\psi}^{\dagger}(1) \hat{b}^{\dagger} \hat{\psi}(0)+ h.c.\right] \,.  
\label{H_JC-Schwinger}
\end{align}
Next we define operators, which measure the charge at the lattice site $x$ i.e.
\begin{subequations}
\begin{align}
\label{eq:Q0_JCM}
\hat{Q}(0) &= \hat{\psi}^{\dagger}(0) \hat{\psi}(0) \, , \\
\label{eq:Q1_JCM}
\hat{Q}(1) &= - \hat{\psi}(1) \hat{\psi}^{\dagger}(1) = \hat{\psi}^{\dagger}(1) \hat{\psi}(1) - 1 \, .
\end{align}
\end{subequations}
We defined $\hat{Q}(x)$ such that the state $\ket{1}=\hat{\psi}^{\dagger}(1)\ket{\mathrm{vac}}_F$ is associated with the absence of any charges  at both sites: $\hat{Q}(0)\ket{1}=\hat{Q}(1)\ket{1}=0 \ket{1}$. On the other hand, the state $\ket{0}=\hat{\psi}^{\dagger}(0)\ket{\mathrm{vac}}_F$ has charge $+1$ at site $0$ and $-1$ at site $1$,  i.e. $\hat{Q}(0)\ket{0}= +1 \ket{0}$ and  $\hat{Q}(1)\ket{0}= -1 \ket{0} $.  
In particular, we will use the notion of such a local charge to discuss local gauge symmetries, which we will introduce in the 
following.
We call a unitary transformation a symmetry, if the Hamiltonian remains invariant with respect to this transformation. These can be space transformations, such as translations and rotations, or transformations between internal degrees of freedom 
such as flavor and color. A transformation which is applied to all degrees of freedom of a system is called a \emph{global transformation}, while a \emph{local transformation} applies only to a finite range of space. 
In particular the JCM is
invariant  under the following two local transformations
\begin{subequations}
\begin{align}
\hat{\psi}(0) &\rightarrow e^{i\theta(0)} \hat{\psi}(0)\, ,\quad \hat{b} \rightarrow e^{+i\theta(0)} \hat{b}  \hspace{-4em} &  \text{(transformation at } x = 0 \text{)}\, , \\
\hat{\psi}(1)  &  \rightarrow e^{i\theta(1)} \hat{\psi}(1)\, ,\quad \hat{b} \rightarrow e^{-i\theta(1)} \hat{b} 
  \hspace{-4em} &  \text{(transformation at } x = 1 \text{)}\, ,
\end{align}
\end{subequations}
{which are parametrized by the angle $\theta(0), \theta(1) \in [0, 2\pi) $.}
These transformations are local since they only involve the space point $0$ or $1$ respectively, and are realized by the unitary operators
\begin{subequations}
\begin{align}
\hat{S}(0) &= e^{i\theta(0) \GJC(0)} \, , \\
\hat{S}(1) &= e^{i\theta(1) \GJC(1)} \, ,
\end{align}
\end{subequations}
with the generators 
\begin{subequations}
\begin{align}\label{eq:GaussJC0}
\GJC(0) &= -\hat{b}^{\dagger} \hat{b} -\hat{Q}(0)  \, , \\
\label{eq:GaussJC1}
\GJC(1) &=  \phantom{-}\hat{b}^{\dagger} \hat{b} - \hat{Q}(1)\, .
\end{align}
\end{subequations}
The unitary operations $S(x)$ leave the Hamiltonian invariant,
\begin{align}
\hat{S}(x) \hat{H}_{JC} \hat{S}^{\dagger}(x) &= \hat{H}_{JC} \quad\forall x
\end{align}
or, equivalently, their generators commute with the Hamiltonian
\begin{equation}
[\GJC(x),\hat{H}_{JC}]=0\quad\forall x \,.
\end{equation}

The local symmetry transformation $\hat{S}(x)$ is called a \emph{ local gauge transformation}, and the commutation relation with the Hamiltonian implies that $\hat{H}_{JC}$ and the symmetry generators $\GJC\left(x\right)$ can be diagonalized simultaneously. Choosing a basis which diagonalizes $\GJC\left(x\right)$ leads to a block-diagonal Hamiltonian. The blocks of the Hamiltonian act on subspaces of the Hilbert space. 
These separate \emph{sectors} are characterized by the eigenvalues of the symmetry generators, and are not mixed by the dynamics generated by $\hat{H}_{JC}$. A \emph{physical} state belongs to a particular sector and does not involve any superposition of different sectors. We formalize this by considering the
eigenvalue problem of \emph{all} generators 
\begin{equation}
\GJC(x)\ket{{\phi}} = q(x) \ket{{\phi}} \quad\forall x \label{eq:GaussLawJC}
\end{equation}
with the eigenvalues $q(x)$, which we call \emph{static charges}. These equations split the Hilbert spaces into different sectors, labeled by sets of static charges. We call the eigenstates of the $\GJC(x)$ operators \emph{physical} states
and the eigenvalue equations \eqref{eq:GaussLawJC}  are denoted as 
\emph{Gauss's laws}. The Gauss's law enforces a \emph{charge superselection rule}, which enforces \emph{physical} states to reside within one respective sector.
Furthermore, the physical states $\ket{\phi}$ are \emph{gauge invariant} i.e. 
they are invariant up to a phase after performing a local gauge transformation, and this phase depends on the static charges
\begin{equation}
\hat{S}(x) \ket{\phi} = e^{i\theta(x) q(x) } \ket{\phi}  \quad \forall x \,.
\end{equation} 
The \emph{local gauge invariance} guarantees the block diagonal structure of the Hamiltonian, which is a direct sum over the different static charge sectors,
\begin{equation}
H = \underset{q}{\bigoplus}H|_{q} \,.
\end{equation}

Besides the local symmetry, the JCM is invariant with respect to the global gauge transformation for fermionic operators $\hat{\psi}(x) \rightarrow e^{i\theta} \hat{\psi}(x)$ leading to the conservation of the total number of fermions 
\begin{equation}\label{eq:JCM-N_F-conservation}
\hat{N}_{F}= \sum_x \hat{\psi}^{\dagger}(x) \hat{\psi}(x)\,,
\end{equation}
i.e. $[\hat{N}_F,\hat{H}_{JC}] =0$. Introducing also the operator of the total number of excitations (the number of photons and the number of atomic excitations) 
\begin{equation}
\hat{N}_\mathrm{ex}= \hat{b}^{\dagger} \hat{b} + \hat{\psi}^{\dagger}(0) \hat{\psi}(0)\,,
\label{eq:N_ex}
\end{equation}
 with  $[\hat{N}_\mathrm{ex},\hat{H}_{JC}] =0$, the generators $\GJC(0)$ and $\GJC(1)$ can be represented in terms of 
the conserving operators
$\hat{N}_\mathrm{ex}$ and $\hat{N}_{F}$ as:  
\begin{subequations}
\begin{align}\label{eq:GaussJC0-N,N_ex}
\GJC(0) &=  -\hat{N}_\mathrm{ex}  \, , \\
\label{eq:GaussJC1-N,N_ex}
\GJC(1) &=  \hat{N}_{F}-1+\hat{N}_\mathrm{ex} \, . 
\end{align}
\end{subequations}

For all physical states one has $\hat{N}_F\ket{\phi} = n_F\ket{\phi}$ with $n_F=1$. In fact, the atom can be only in the internal states $\ket{x}=\hat{\psi}^{\dagger}(x)\ket{\mathrm{vac}}_F$ containing one fermion
 in the Schwinger representation.
Consequently the \emph{total charge} operator
\begin{align}
\hat{Q} = \sum_x \hat{Q}(x)=\hat{N}_{F}-1 
\label{eq:Q_JCM}
\end{align}
has zero eigenvalues for physical states with $n_F=1$, giving
\begin{align}
\sum_{x} q(x) = 0\,.
\end{align}

In order to determine the spectrum of the Jaynes-Cummings Hamiltonian we
use the property $q(0) = -q(1)$, which implies that the physical sectors can
be denoted by one positive integer $q \equiv q(1)$. The eigenstates of the two Gauss's laws 
\eqref{eq:GaussLawJC} are

\begin{subequations}
\begin{align}
\label{1,q}
\ket{1,q} &= \hat{\psi}^{\dagger}(1)\ket{\mathrm{vac}}_F\otimes\ket{q}_B \, ,\\
\label{0,q}
\ket{0,q} &=  \hat{\psi}^{\dagger}(0)\ket{\mathrm{vac}}_F\otimes\ket{q-1}_B \, ,
\end{align}
\end{subequations}
where $\ket{\mathrm{vac}}_F$ is the empty fermionic Fock state and $\ket{q}_B$ is a bosonic state with $q$ particles.
Note that the static charge $q \equiv q(1)$ represents the number of excitations (photons or excited atoms) in the system.
Each state in the sector $q$ can be expanded as
\begin{align}
\ket{\phi_q} = \alpha_q \ket{1,q}+ \beta_q \ket{0,q} \,.\label{eq:eigenstate_JCM}
\end{align}
Within the sector with $q\ge1$ the Hamiltonian matrix using the basis $\{\ket{1,q}, \ket{0,q}\}$ will take the form 
\begin{equation}
H|_{q} = \begin{pmatrix}
\omega_cq - \frac{\omega_a }{2} & \tfrac{ \Omega_0}{2} \sqrt{q}\\
\tfrac{ \Omega_0}{2} \sqrt{q} &  \omega_c  (q-1) + \frac{\omega_a }{2} 
\end{pmatrix}
\,.\label{eq:eigen-value-equation-JCM-Abelian}
\end{equation}
The eigenvalues of the Hamiltonian \eqref{eq:eigen-value-equation-JCM-Abelian} 
\begin{align}
E_{\pm,q} = \omega_c (q - \tfrac{1}{2} ) \pm \frac{1}{2} \sqrt{\Delta^2 + \Omega_0^2q}\,
\label{eq:E_pm,q-JCM}
\end{align}
make the Jaynes-Cummings ladder by taking different sectors $q$, where we introduced the detuning $\Delta = \omega_c - \omega_a$. The corresponding eigenvectors are
\begin{subequations}
\begin{align}
\label{eq.ket_q+}
\ket{q,+} &= \cos(\alpha_q/2) \ket{1,q} + \sin(\alpha_q/2) \ket{0,q} \, , \\ 
\label{eq.ket_q-}
\ket{q,-} &= \sin(\alpha_q/2) \ket{1,q} - \cos(\alpha_q/2) \ket{0,q}
\end{align} 
\end{subequations}
with $\tan(\alpha) = \Omega_0 \sqrt{q}/ \Delta $.
For $q=0$ there is only one eigenstate of the Gauss's law $\ket{1,q} = \hat{\psi}^{\dagger}(1)\ket{\mathrm{vac}}_F\otimes\ket{q}_B$ characterized by zero eigenvalue $q$. The state $\ket{1,q} $ is an eigenvector of the Hamiltonian with an eigenenergy $-\omega_c /2$. 

\section{Lattice gauge theories: the case of compact U(1) \label{sec:3}}
In order to generalize the one link model of the previous section we consider
a one dimensional open lattice (chain) with an  even number of sites $\mathcal{N}=2\mathcal{M}+2$. The sites $x$ are labeled from $0$ to $\mathcal{N}-1=2\mathcal{M}+1$. Each site may host at most one fermion, created by $\psi^{\dagger}\left(x\right)$. 
In order to introduce the concept of particles, anti-particles and charge we consider the mass Hamiltonian of this one-dimensional system 
\begin{align}
H_M = M \sum_x (-1)^x \hat{\psi}^{\dagger}(x) \hat{\psi}(x) \, , \label{eq:Staggered}
\end{align}
where the alternating sign manifests that the fermions are \emph{staggered} \cite{Susskind1977}:
an occupied even site corresponds to a particle and an empty odd site corresponds to an anti-particle.  Note that the staggered mass is given by $M=\omega_a / 2$ in the Jaynes-Cummings Hamiltonian  \eqref{H_JC-Schwinger}. 
Respectively, we define the local charge operators as 
\begin{equation}
\hat{Q}(x) = \hat{\psi}^{\dagger}(x) \hat{\psi} (x) - t(x) 
\label{eq:Charge}
\end{equation}
where $t\left(x\right)=0(1)$ for even (odd) site $x$. The definition of the local charge operators \eqref{eq:Charge} is analogous to Eqs. \eqref{eq:Q0_JCM} and \eqref{eq:Q1_JCM} for the two site JCM.

The mass Hamiltonian conserves the total number of fermions, and we shall fix it to $N_F = \mathcal{M} + 1$ i.e. \emph{half filling}. 
We identify the groundstate of the mass Hamiltonian \eqref{eq:Staggered} at half-filling with the \emph{Dirac Sea}. In this groundstate the odd sites are occupied with one particle, whereas the even sites are empty - corresponding to a configuration in which no particles or anti-particles are present.  

\subsection{Compact U(1) on a single link}\label{Compact U(1)}

\begin{figure}\centering{\includegraphics[width=\columnwidth]{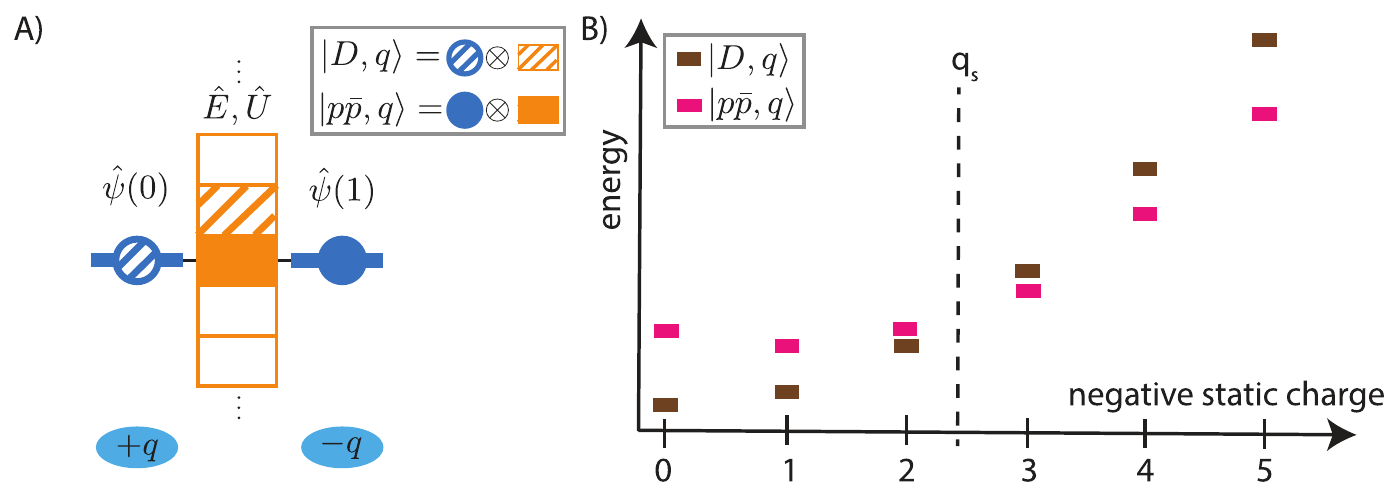}
\caption{\label{Fig:cQED_One_Link} Visualization of a single link and spectrum of compact QED for $g = M = 1$. A) The left lattice site associated with $\hat{\psi}(0)$ depicts the state $\ket{D,q}$, the right site associated with $\hat{\psi}(1)$ shows the $\ket{p\bar{p},q}$. The orange ladder illustrates the Hilbert space of the link associated with the electric field $\hat{E}$ and gauge connection $\hat{U}$. B) Energy of the Dirac vacuum $\ket{D,q}$ and particle-anti-particle state $\ket{p\bar{p},q}$ determined by the diagonal of \eqref{eq:eigen-value-equation-U(1)link}. The state $\ket{p\bar{p},q}$ becomes stable above a negative static charge of $q_s$   }
}
\end{figure}

In this subsection we will discuss a model consisting out of two sites and one link connecting these sites. This model will be a minimal instance of a  U(1) lattice gauge theory. As in the previous section the fermions reside on the sites. 
We consider states which are dynamically connected to the Dirac sea $\ket{D} = \hat{\psi}^\dag(1) \ket{\mathrm{vac}}_F$ with zero charge. 
The occupation of the site $0$ can be understood as a particle anti-particle state $\ket{p\bar{p}} = \hat{\psi}^\dag(0) \ket{\mathrm{vac}}_F$ with fermionic charges $Q(1) = -1$ and $Q(0)=1$.

In the JCM considered in Sec.~\ref{sec:2}, the Hilbert space of the link between the two lattice sites 
comprises a boson mode described by raising and lowering operators $\hat{b}^{\dagger} $ and $ \hat{b} $.
In the U(1) gauge theory the Hilbert space of the link is different and represents the Hilbert space of a \emph{particle on a ring}. The position coordinate $\varphi$ of a particle on a ring is an angle, residing in a compact configuration space. On  the link, the angular position operator 
  $\hat{\varphi}$ defines a continuous basis through its eigenstates  $\ket{\varphi}_G$, which span the Hilbert space of the link. The Hilbert space of the line will play the role of a gauge degree of freedom, hence the subscript $G$.  
Moreover we use the angle operator $\hat{\varphi}$ to define the unitary operator 
\begin{align}
\hat{U} = e^{i\hat{\varphi}} \, ,
\label{U}
\end{align}
which is invariant with respect to shifting the phase operator $\hat{\varphi}$ by $2\pi$. 
The operator $\hat{U}$ 
induces a representation of the U(1) group on the Hilbert space of a particle i.e. 
\begin{align}
\hat{U} \ket{\varphi}_G = e^{i\varphi} \ket{\varphi}_G\,. 
\label{U-eigenstate-eq}
\end{align}
The operator $\hat{U}$ is called the \emph{gauge connection} or \emph{group element operator}. The conjugate momentum of the angle operator will be the angular momentum operator $\hat{E}$, which we call the electric field. It is given by $-\mathrm{i}\partial_{\varphi}$ in the position representation of the angular motion.  The angular momentum operator has a non-bounded integer spectrum
\begin{align}
\hat{E} \ket{j}_G = j \ket{j}_G \, ,
\end{align} 
where $j \in \mathbb{Z}$.
The canonical commutation relation 
\begin{equation}
[\hat{\varphi} ,\hat{E}  ] \equiv [\varphi , -\mathrm{i}\partial_{\varphi}  ]=i \,
\label{[varphi,E]}
\end{equation}
 leads to
\begin{align}
[\hat{E},\hat{U}] = \hat{U} \, 
\end{align}
- that is, $ \hat{U} $ raises the electric field by one unit:
\begin{align}
\hat{U} \ket{j}_G = \ket{j+1}_G \, 
\end{align} and the conjugate operator $ \hat{U}^{\dagger} $ lowers the electric field.

After introducing link variables we are able to formulate a one link gauge theory in analogy to the JCM. The Hamiltonian 
takes the following form:
\begin{align}
\hat{H}_{1}= \frac{g^2}{2} \hat{E}^2  + M \left[\hat{\psi}^\dag(0) \hat{\psi}(0) - \hat{\psi}^\dag(1) \hat{\psi}(1)\right]+ \epsilon \left[ \hat{\psi}^{\dagger}(0) \hat{U} \hat{\psi}(1) + h.c \right] \,.
\label{H_U1}
\end{align}
In this model, the cavity energy $\omega_c \hat{b}^{\dagger} \hat{b}$ featured in the Jaynes-Cummings Hamiltonian \eqref{H_JC-Schwinger} is substituted by the operator $\hat{E}^2$ corresponding to the kinetic energy of a rotating particle. The second term in Eq.~\eqref{H_U1} is the staggered mass term and the last term describes the creation of the particle-antiparticle pair out of the Dirac vacuum associated with the \textit{increase} of the electric field on the connecting link and vice-versa. 


The Hamiltonian is invariant under to the following local transformations:
\begin{subequations}
\begin{align}
\label{psi->psi'}
\hat{\psi}(0) &\rightarrow e^{i\theta(0)} \hat{\psi}(0)\, ,\quad \hat{U}(0) \rightarrow e^{+i\theta(0)} \hat{U}(0)  \hspace{-2em} &  \text{(transformation at } x = 0 \text{)}\, , \\
\hat{\psi}(1)  &  \rightarrow e^{i\theta(1)} \hat{\psi}(1)\, ,\quad \hat{U}(0) \rightarrow e^{-i\theta(1)} \hat{U}(0) 
  \hspace{-2em} &  \text{(transformation at } x = 1 \text{)}\, ,
\end{align}
\end{subequations}
where the phase operator transformed as 
\begin{subequations}
\begin{align}
\hat{\varphi}  \rightarrow \hat{\varphi}  + \theta(0) \,  & \quad \text{(transformation at } x = 0 \text{)}\, , 
\label{varphi(0)-gauge-transform}
\\
\hat{\varphi}  \rightarrow \hat{\varphi}  - \theta(1) \,  & \quad \text{(transformation at } x = 1 \text{)}\, .
\label{varphi(1)-gauge-transform} 
\end{align}
\end{subequations}
The transformation of the phase resembles the form of the gauge transformation of the vector potential in the continuum. Thus one identifies the phase operator $\hat{\varphi}$ 
with a \emph{compact vector potential}  $\hat{A}=\hat{\varphi}$  obeying the canonical commutation relation \eqref{[varphi,E]}, i.e. $[\hat{A} ,\hat{E} ]=\mathrm{i} $.
This is a very natural choice of the vector potential, as, using the solid state terminology, the unitary operator $\hat{U} = e^{i\hat{A}} $  entering the Hamiltonian \eqref{H_U1} can be interpreted as a Peierls's phase factor \cite{Peierls33ZPh,Luttinger51PR,Hofstadter76PRB} for fermion transition between the lattice sites $1$ and $0$. The vector potential $\hat{A}$ acts then in the Hilbert space of the quantum gauge field living on the link between the two lattice sites.  
The local transformations are implemented by the operators
\begin{subequations}
\begin{align}
\hat{S}(0) &= e^{i\theta(0) \hat{G}(0)}  \, ,\\
\hat{S}(1) &= e^{i\theta(1) \hat{G}(1)} \, ,
\end{align}
\end{subequations}
with the generators 
\begin{subequations}
\begin{align}\label{Eq:GaussJC}
\hat{G}(0) &= \phantom{-}\hat{E} - \hat{Q}(0) \, ,\\
\hat{G}(1) &=  -\hat{E}  - \hat{Q}(1) \, ,
\end{align}
\end{subequations}
where $\hat{Q}(1)$ and $\hat{Q}(0)$ are local charge operators given by Eq. \eqref{eq:Charge}. The unitary transformation $\hat{S}(x)$ leaves the Hamiltonian invariant
\begin{align}
\hat{S}(x) \hat{H}_{1} \hat{S}^{\dagger}(x) &= \hat{H}_{1} \quad\forall x
\end{align}
or, in terms of the generators,
\begin{equation}
\left[\hat{G}(x),\hat{H}_{1}\right]=0\quad\forall x \,.
\end{equation}
Again we define physical states as the eigenstates of the Gauss's law operators
\begin{equation}
\hat{G}(x)\ket{{\phi}} = q(x) \ket{{\phi}} \quad\forall x \,.\label{eq:GaussLaw}
\end{equation}
Where the last equation immediately implies the splitting of the Hilbert space into charge sectors. As in the Jaynes-Cummings case, 
half-filling  of fermions  ($n_F=1$) guarantees that $\sum_x q\left(x\right)=0$.  Thus
one can write $q \equiv q\left(0\right)$ and $ q\left(1\right)=-q$. The charge sector $q$ may be spanned by two states, as in Eqs. (\ref{1,q}) and (\ref{1,q}) for the JCM:
\begin{subequations}
\begin{align}
\ket{D,q} &= \hat{\psi}^{\dagger}(1)\ket{\mathrm{vac}}_F\otimes\ket{q}_G \, ,\\
\ket{p\bar{p},q} &=  \hat{\psi}^{\dagger}(0)\ket{\mathrm{vac}}_F\otimes\ket{q+1}_G \, ,
\end{align}
\end{subequations}
where $\ket{D,q}$ corresponds to a state with no particles and anti-particles (Dirac sea), in which the static charges are the only source of electric field, while 
$\ket{p\bar{p},q}$ is a meson-like state with a particle anti-particle pair, serving as the source of an electric flux tube on top of the one originating from the static charges.

A general state of the $q$ sector is then spanned by
\begin{align}
\ket{\phi_q} = \alpha_q \ket{D,q}+ \beta_q \ket{p\bar{p},q}
\end{align}
and the matrix block of the Hamiltonian, which corresponds to the $q$ sector, takes the form
\begin{equation}
H_1|_q = \left(\begin{array}{cc}
  \frac{g^2}{2}q^2-M& \epsilon\\
\epsilon & \frac{g^2}{2} (q+1)^2+M
\end{array}\right) \,.\label{eq:eigen-value-equation-U(1)link}
\end{equation}
The off-diagonal elements of the previous matrix are independent of $q$, whereas the off-diagonal elements of the Jaynes-Cummings Hamiltonian \eqref{eq:eigen-value-equation-JCM-Abelian} depend on the sector. 

In the limit of small interaction (negligible off diagonal terms), the self energy is approximately given by the diagonal terms. If the static charges have negative values, $q<0$,  they imply an external electric field which is of opposite sign to the one created by a particle-antiparticle pair. Hence the creation of the state $\ket{p\bar{p},q}$ actually lowers the electric field amplitude in the system. For strong enough negative values of the static charge $q$, this effect even makes the state $\ket{p\bar{p},q}$ of lower energy than the state $\ket{D,q}$. From equating the diagonal terms, we find that this happens for 
\begin{equation}\label{Eq:SchwingerPairSimple}
q \approx -\left(\frac{2M}{g^2}+\frac{1}{2}\right) \, ,
\end{equation}
which can be understood in terms of a two level system. 
For a single spin in an external magnetic field with the Hamiltonian $H_1|_q  = \frac{g^2}{2} \left[\left(q+\frac{1}{2}\right)^2+\frac{1}{2}\right] \mathbbm{1} +(2M+2q+1)\hat{s}_z + 2\epsilon \hat{s}_x$, where $\hat{s}_{x,z}$ are spin-half operators. The $z$-component of the magnetic field inverts sign at \eqref{Eq:SchwingerPairSimple}, hence inverting the state of minimal energy.
The one-link cQED is the minimal model allowing to demonstrate the instability of the fermionic vacuum in the presence of strong electric fields, see also recent work \cite{ilderton2020renormalisation, ilderton2020toy}. The effect is reminiscent of the widely studied Schwinger pair production in extended systems \cite{Schwinger1951, Kasper2014, Kasper2016}. 

\subsection{An Abelian lattice gauge theory: (1+1) dimensional cQED} \label{(1+1) cQED}
\begin{figure}\centering{\includegraphics[width=0.8\columnwidth]{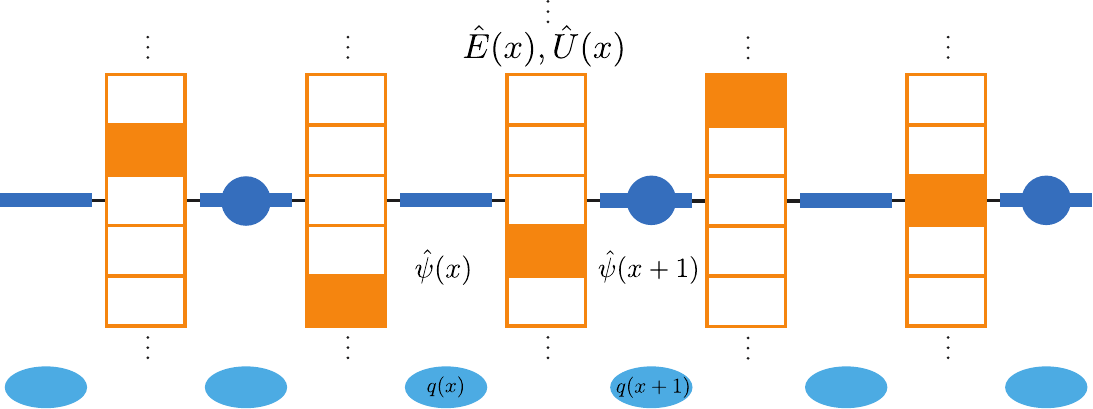}
\centering
\caption{\label{Fig:cQED} Visualization of compact QED. The system consists of matter sites (blue), which are linked by gauge fields (yellow). The fermionic 
operators $\hat{\psi}(x)$ act on the sites, whereas the electric fields $\hat{E}(x)$ and $\hat{U}(x)$ act on the links. The labels $q(x)$ denote the charge sectors.}
}
\end{figure}

Next, we extend the one-link model of the previous section and
include several vertices and links, while staying one-dimensional.
The resulting model is a one dimensional lattice gauge theory: compact quantum electrodynamics (cQED) in one space dimension with open boundary conditions. The lattice has an even number of sites $\mathcal{N}=2\mathcal{M}+2$, which are labeled by $x=0,...,\mathcal{N}-1$. 
 On each link we have the Hilbert space of a particle on a ring, with the gauge connection operators $\hat{U}(x) = e^{i\hat{\varphi}(x)} $ and  electric fields $\hat{E}(x)$. 
The Hamiltonian for $(1+1)$ cQED is
\begin{align}
\hat{H}_{cQED}=\frac{g^2}{2} \sum_{x} \hat{E}^2(x) 
+M \sum_x (-1)^x \hat{\psi}^\dag(x)  \hat{\psi}(x) 
+\epsilon \sum_x  \left[ \hat{\psi}^{\dagger}(x)  \hat{U}(x) \hat{\psi}(x+1) +h.c.\right]\,,
\label{eq:c-QED}
\end{align}
where the summation over $x$ covers all the lattice sites ($x=0,...,\mathcal{N}-1$) in the second term, whereas the sum is over all  links ($x=0,...,\mathcal{N}-2$) in the first and the third terms. {For a graphical illustration of the Hamiltonian Eq.~\eqref{eq:c-QED} see Fig.~\ref{Fig:cQED}}. The first term is the electric field energy with $g$ the coupling constant. The second term is the staggered mass term and the last
term is the coupling between the gauge field and the matter degrees of freedom.
In the single link case ($\mathcal{N}=2$) Eq. \eqref{eq:c-QED} reduces to the Hamiltonian given by Eq.~\eqref{H_U1} subject to replacement $\hat{E}^2(0) \rightarrow \hat{E}^2$ and $ \hat{U} \left(0\right) \rightarrow \hat{U}$.
$H_{cQED}$ is symmetric under the local (gauge) transformations
\begin{align}
&\hat{\psi}(x) \rightarrow e^{i\theta(x) }\hat{\psi}(x) \, , \,
\hat{U}(x)  \rightarrow e^{i\theta(x)} \hat{U}(x) \, , \,
 \hat{U}(x-1) \rightarrow e^{-i \theta(x)} \hat{U}(x-1) \, \label{eq:GaugeTrafoBulk}
\end{align}
for $0<x<\mathcal{N}-1$, and at the boundaries
\begin{subequations}
\begin{align}
\hat{\psi}(0) &\rightarrow e^{i\theta(0) }\hat{\psi}(0) \, , \,
\hat{U}(0)  \rightarrow e^{i\theta(0)} \hat{U}(0)  \, ,  \\
\hat{\psi}(\mathcal{N}-1) &\rightarrow e^{i\theta(\mathcal{N}-1) }\hat{\psi}(\mathcal{N}-1) \, , \,
\hat{U}(\mathcal{N}-2) \rightarrow e^{-i \theta(\mathcal{N}-1)} \hat{U}(\mathcal{N}-2) \, .
\end{align}
\label{eq:GaugeTrafoBoundary}
\end{subequations}
{The local gauge transformations of the matter field $\hat{\psi}(x)$ and the link $\hat{U}(x)$ are different in the bulk \eqref{eq:GaugeTrafoBulk} and at the boundaries Eq.~\eqref{eq:GaugeTrafoBoundary}. In particular, the gauge transformations are different at the boundaries since matter cannot move across the boundary and there is no electric flux out of the system. Further, matter not moving out of the system ensures that the total charge is conserved.} This local gauge transformation is represented by the unitary transformation
\begin{align}
\hat{S}(x) = e^{i \theta(x) \hat{G}(x)}
\end{align}
with the generators
\begin{align}
\hat{G}(x)  &= \hat{E}(x) - \hat{E}(x-1) - \hat{Q}(x) 
\end{align}
for $0<x<\mathcal{N}-1$, and at the boundaries
\begin{subequations}
\begin{align}
\hat{G}(0) &= \hat{E}(0) - \hat{Q}(0) \, , \\
\hat{G}(\mathcal{N}-1) &= -\hat{E}(\mathcal{N}-2) - \hat{Q}(\mathcal{N}-1) \, ,\end{align}
\end{subequations}
using the local charges $Q(x)$  defined in \eqref{eq:Charge}.
The symmetry generators fulfill
\begin{equation}
[\hat{G}(x),\hat{H}_{cQED}]=0\quad\forall x
\end{equation}
and in a complete analogy to the two previous cases, we define the gauge invariant, or physical states $\ket{\phi}$, as
\begin{equation}
\hat{G}(x) \ket{\phi} = q(x) \ket{\phi} \,.
\end{equation}

The Gauss's laws give rise to superselection sectors labeled by the configuration of static charges $\left\{q(x)\right\}$. In particular the Gauss's laws of the bulk are
\begin{equation}
[ \hat{E}(x) - \hat{E}(x-1)] \ket{\phi} = 
[\hat{Q}(x)  +q(x) ] \ket{\phi}
\end{equation}
- the divergence of electric fields at a vertex equals the sum of dynamical and static charges there, as expected in a Gauss's law. Since the global and local U(1) symmetry are related, the conservation of the total number of fermions is connected to the global charge of the system:
\begin{equation}
\hat{Q}=\sum_x \hat{Q}(x)=\hat{N}_F - (\mathcal{M}+1)\,. 
\label{Q_1+1cQED}
\end{equation}
Thus the operators of the total particle number $\hat{N}_F$  and the  global charge 
$\hat{Q}$ differ only by a constant shift of $\mathcal{M}+1=\mathcal{N}/2$ representing the number of odd sites in the lattice. The latter $\mathcal{M}+1$ equals to the unity in Eq.~\eqref{eq:Q_JCM} for the JCM.
As in the Jaynes-Cummings case, adding all the Gauss's laws leads to a constraint for the static charges:
\begin{equation}
\underset{x}{\sum}q(x) \ket{\phi}= 
\underset{x}{\sum}\hat{G}(x)  \ket{\phi}=
-\underset{x}{\sum} \hat{Q}(x)  \ket{\phi}=0 \,,
\end{equation}
where the last equality ($\ldots=0$) holds 
since we consider fermionic half-filling ($n_F=\mathcal{M}+1$) configurations in the lattice, which includes also the Dirac sea state.

\subsection{Hilbert space dimension of (1+1) dimensional cQED}
The Hilbert space of the $(1+1)$ dimensional lattice gauge theory is a tensor 
product of the Fock space of the matter sites, which is finite dimensional for a finite system, and the Hilbert spaces on the links, which are infinite dimensional. However, the gauge symmetry restricts the physical Hilbert space into a subspace of the infinite many-body Hilbert space. It is a union of all possible static charge sectors which are labeled by all possible integers $\left\{q(x) \right\}$ satisfying the condition $\sum_x q(x) = 0$. A sector denoted by a specific static charge configuration $\left\{q(x) \right\}$ cannot be left with the gauge invariant dynamics governed by the Hamiltonian $\hat{H}_{cQED}$ given by \eqref{eq:c-QED}. For a finite chain, the dimension of one sector will be finite, like in the U(1) single link model considered in Sec.\ref{Compact U(1)}. In the following we derive an upper bound for the dimension of one
sector for the lattice of $\mathcal{N}=2\mathcal{M}+2$ sites. 

The Gauss's law on the first vertex is
\begin{equation}
\hat{E}(0) \ket{\phi \left\{q(x) \right\} }
= \left[q(0) + \hat{Q}(0) \right]\left|\phi  \left\{q(x) \right\}\right\rangle \,.
\end{equation}
Because the eigenvalues of  $\hat{Q}(0)$ can only be $0$ or  $1$, we will only have two possibilities for $\hat{E}(0)$, \textit{viz.} $\hat{E}(0) = q(0)$ and $\hat{E}(0) = q(0)+1$.
The Gauss's law at the next link is
\begin{equation}
\hat{E}(1) \ket{ \phi \left\{q(x) \right\} } 
= \left[\hat{E}(0) + q(1) + \hat{Q}(1)\right] \ket{\phi \left\{q(x) \right\}} \,.
\end{equation}
Since $Q(1)  = 0,-1$ we deduce that we have three options for $E(1) $: 
$q(0) + q(1) -1$, $q(0) +q(1)$ and $q(0)+q(1)+1$. Similarly, the next link will require a four dimensional Hilbert space, spanned by the electric field values $E(2)$ equal to $q(0)+q(1)+q(1)-1$, $q(0)+q(1)+q(1)$, $q(0)+q(1)+q(1)+1$ and $q(0)+q(1)+q(1)+2$. Moving on recursively, we obtain that the allowed values of $E(n)$ are $\sum_{x=0}^n q(x) - \frac{n}{2},...,\sum_{x=0}^n q(x) + \frac{n}{2} +1$ if  the site $n$ is even, and $\sum_{x=0}^n q(x) - \frac{n+1}{2},...,\sum_{x=0}^n q(x) + \frac{n+1}{2}$ if $n$ is odd.

Similarly we can start from  the Gauss's law at position $\mathcal{N} -1 $ and move toward $x = 0$ repeating the recursive argument. Hence the Hilbert space dimension of the link at position $n$ and at position $2\mathcal{M}-n$ is the same i.e. 
$\dim\left(n\right) = \dim\left(2\mathcal{M}-n\right)$
and, for $0\leq n \leq \mathcal{M}$ the dimension is given by
\begin{equation}
\dim\left(n\right) = n+2 \,.
\end{equation}
Therefore the  maximal Hilbert space dimension of all the links is given by
\begin{equation}
\mathcal{D}_{\text{Gauge}}  \left(\mathcal{M}\right)\equiv  \left(\prod_{n=0}^{\mathcal{M}-1}
\left(n+2\right)\right)^2\left(\mathcal{M}+2\right)=\left(\mathcal{M}+1\right)!\left(\mathcal{M}+2\right)! \,, 
\end{equation}
which is bounded for a finite system size. 

The fermionic Fock space dimension reduces in the half filling case to  $\binom{2\mathcal{M}+2}{\mathcal{M}+1}$ - number of possibilities to distribute $\mathcal{M}+1$ on $2\mathcal{M}+2$ sites. Hence,  
\begin{equation}
\mathcal{D}\left(\mathcal{M}\right)< \left(\mathcal{M}+2\right)\left(2\mathcal{M}+2\right)!=\mathcal{N}!\left(\mathcal{N}/2+1\right)
\end{equation} 
is an upper bound for Hilbert space describing the physics of (1+1) cQED  with open boundary conditions. This bound is independent of the charge sector. In particular it is not a strict 
bound since gauge invariance has not been fully exploited. In one space dimension with open boundaries one can solve the Gauss's law explicitly, but this gives rise to non-local interactions \cite{Hamer1997, Bringoltz2009,Banuls2013,Martinez2016,Emonts2020}. The above upper bound denotes dimensions which will be required for building a quantum simulator which does not eliminate the Gauss's law and involves only local interaction terms.

\section{Non-Abelian Generalization \label{sec:4}}
In the following sections we will generalize from Abelian gauge theories to 
non-Abelian gauge theories. Whereas the elements of Abelian groups commute, this is not the general case of those of a non-Abelian groups. Famous examples are $SO(3)$ - the group of rotations in three dimensional space, or SU(2), the symmetry group of spins. We will focus on SU(2) and use a formalism similar to \cite{Zohar2015} for discussing Hamiltonian lattice gauge theories as first introduced by Kogut and Susskind \cite{Kogut1975}.

\subsection{SU(2) gauge degree of freedom: the rigid rotor}

We begin by reviewing relevant properties of SU(2) and use this to formulate 
a non-Abelian one link model with local gauge invariance. We first introduce the relevant Hilbert spaces on the links and on the vertices. The Hilbert space, which will be used on the
link, is the same as the quantum rigid rotor, whereas the Hilbert space of vertices 
will be built by fermionic particles. Most importantly the link Hilbert space as
well as the fermionic Hilbert space on the vertices will 
support representations of SU(2). 

We start by introducing the group SU(2) as the rotation of an object 
parametrized by two angles $\alpha$ and $\beta$ e.g. a particle on a sphere, a spin or a diatomic molecule. However, the group SU(2) also appears by studying the rotation of a rigid 
object. In classical mechanics \cite{Goldstein2002} as well as in quantum 
mechanics the rigid body is determined by three angles - the Euler angles - $\alpha$, $\beta$, 
$\gamma$.  While we shall use Euler angles to characterize the rigid body there are other ways to parametrize any group element $g \in SU(2)$, which will involve three independent parameters $\theta_a\left(g\right)$.

A general spin rotation of a state is implemented by the unitary operator  
\begin{align} \label{Eq:TransformSuTwo}
\hat{S}_g&= e^{i \theta_a\left(g\right) \hat{J}_a}\,,
\end{align}
where a summation is implied over a Cartesian index $a$ repeated twice. Here $\theta_a\left(g\right)$ are the group parameters, uniquely identifying one particular rotation $g \in SU(2)$, and the
operators $J_a$ (with $a=x,y,z$) are the three \emph{rotation generators}, also known as the angular momentum operators. These operators do not commute, but rather satisfy the SU(2) Lie algebra
\begin{align}
[\hat{J}_a, \hat{J}_b]&= i \epsilon_{abc} \hat{J}_c \, ,
\end{align}
where $\epsilon_{abc}$ is the Levi-Civita symbol.
From these three operators one can build the \emph{total angular momentum} or \emph{quadratic Casimir operator}
\begin{equation}
\hat{\mathbf{J}}^2 = \hat{J}_a \hat{J}_a = \hat{J}_x^2 + \hat{J}_y^2 + \hat{J}_z^2 \, .
\end{equation}
It commutes with all the generators
\begin{equation}
\left[\hat{\mathbf{J}}^2,\hat{J}_a\right]=0 \, 
\end{equation}
and thus one may choose one particular generator - usually $J_z$, and combine it with $\mathbf{J}^2$ to a 
 \emph{maximally mutually commuting set} of operators, which can be simultaneously diagonalized. The quantum numbers given by the eigenvalues of these operators are used as quantum numbers, labeling quantum states with respect to their transformation properties. Those two operators define the \emph{angular momentum states} $\ket{jm}$ as 
\begin{align}
\hat{\mathbf{J}}^2\ket{jm} &= j(j+1)\ket{jm} \, ,\\
\hat{J}_z\ket{jm} &=m\ket{jm}
\end{align}
with  $j = \frac{1}{2} n$, $n\in \mathbb{N}$ and $m = -j,-j+1 \cdots, j$.
The 
quantum number $j$ denotes the irreducible representation of SU(2), which contains $2j+1$ states with different $m$ values. 

General rotations mix the different states within a representation, but not different representations: the rotation operators are block diagonal in the representations.
Therefore the generators may be expressed as a direct sum in the representations,
\begin{equation}
\hat{J}_a = \underset{j}{\bigoplus}(T^{j}_a)_{mn}\ket{jm}\!\bra{jn} \,,
\end{equation}
where summations are implied over the quantum numbers $m$ and $n$ repeated twice. For each irreducible representation $j$, $T^j_a$ is a set of matrices with dimension $\dim\left(j\right)\equiv2j+1$ satisfying the algebra of the group
\begin{equation}
\left[T^j_a,T^j_b\right]=i\epsilon_{abc}T^j_c \,.
\end{equation}
In particular for $j=1/2$, the fundamental representation, the matrices $T^j_a$  are given by
\begin{align}
(T^{1/2}_a)_{mn}&= \frac{1}{2}\sigma_a \, ,
\end{align}
where $\sigma_a$ denotes the three Pauli matrices.
The $j$ irreducible representation of a group element $g$ is given by the unitary matrix
\begin{equation}
D^j\left(g\right) = e^{i\theta_a\left(g\right)T_a^j} \, ,
\end{equation}
which are called \emph{Wigner matrices}. In terms of the Euler angles $\alpha$, $\beta$ and $\gamma$  the Wigner matrices read \cite{Wigner1959}
\begin{equation}
D^j\left(\alpha,\beta,\gamma\right) = e^{-i\alpha T^j_z}e^{-i\beta T^j_y}e^{-i\gamma T^j_z} \,.
\end{equation}
A general rotation operator may therefore be written in the block diagonal form
\begin{equation}
\hat{S}\left(\alpha,\beta,\gamma\right)
= \underset{j}{\bigoplus} D^j_{m'm}\left(\alpha,\beta,\gamma\right)
\left|jm'\right\rangle\left\langle jm\right| \,
\end{equation} 
implying that 
\begin{equation}
\hat{S}(\alpha, \beta, \gamma)\ket{jm}= D^j_{m'm}(\alpha, \beta, \gamma)\ket{jm'}
\end{equation}


In the case of U(1) we considered a Hilbert space parametrized by one angle - the
coordinate for a particle on a ring. In the previous section we discussed rotations of a particle parametrized by two angles. Next, we consider a Hilbert space parametrized by 
three Euler-angles $\ket{\alpha, \beta, \gamma}$ - the configuration (coordinate) space of the quantum mechanical rigid rotor \cite{Landau1981,Weinberg2015}. Let us consider the rotation of an isotropic rigid rotor, whose Hamiltonian is 
\begin{equation}
\hat{H} = \frac{\hat{\mathbf{J}}^2}{2I} \, ,
\end{equation}
where $\hat{\mathbf{J}}^2$ is the total angular momentum. Since we consider
an isotropic rigid body, the moment of inertia is the same for all axis and 
denoted by $I$.

The rotation may be described in the inertial laboratory or \emph{space frame}, but as we are dealing with a rigid body, we may also choose another frame of reference fixed to the body - for example, with its origin at its center of mass - the \emph{body frame}. {The two frames are related by a rotation \cite{Landau1981,Weinberg2015}.}
{Measuring the components of (pseudo) vector operators in two frames will give rise to different results. One thus has to introduce different operators in both frames, which are related by a unitary operation \cite{Weinberg2015}. The
\emph{two} sets of angular momentum operators fulfill in the space frame,}
\begin{equation}
\left[\hat{R}_a,\hat{R}_b\right]=i\epsilon_{abc}\hat{R}_c \, ,
\end{equation}
the SU(2) algebra, and in the body frame
\begin{equation}
\left[\hat{L}_a,\hat{L}_b\right]=-i\epsilon_{abc}\hat{L}_c
\end{equation}
with an opposite sign in the algebra which results from rotating the angular momentum operator from the space frame to the body frame \cite{Landau1981,Weinberg2015}.
{Since one may measure the angular momentum in one frame without
disturbing the measuring angular momentum in the other frame the generators of space and body rotations must commute}
\begin{equation}
\left[\hat{R}_a,\hat{L}_b\right]=0 \, .
\label{[R_a,L_b]}
\end{equation}

On the other hand, the two sets of rotations are not fully independent: the space and body frames are related by a rotation, and hence {a measurement of the} total angular momentum {will give rise to the same result in both non-independent } frames \cite{Weinberg2015}. 
We define a total angular momentum operator, also called Casimir operator,  as
\begin{equation}
\hat{\mathbf{J}}^2\equiv\hat{\mathbf{R}}^2 = \hat{\mathbf{L}}^2   \, ,
\label{JRL}
\end{equation}
which is the same in both frames. 

Again we choose a maximal set of mutually commuting operators: the total angular momentum $\hat{\mathbf{J}}^2$ shared by both frames and the $z$ components i.e. $\hat{L}_z$ and $\hat{R}_z$. This allows to span the Hilbert space of the rigid rotor by angular momentum or \emph{representation basis} state $\ket{jmn}$, defined by
\begin{subequations}
\begin{align}
\hat{\mathbf{J}}^2\ket{jmn} &= j\left(j+1\right)\ket{jmn} \, , \\
\hat{L}_z\ket{jmn} &= m\ket{jmn} \, ,\\
\hat{R}_z\ket{jmn} &= n\ket{jmn} \, .
\end{align} \label{eq:DefinitionOfJKM}
\end{subequations}
The eigenstates of the rigid body rotor $\ket{jmn} $ are characterized by three quantum numbers $j$, $m$ and $n$, where $m$ and $n$ are associated with two projections of the total angular momentum $j$. In contrast the angular motion of a quantum particle in a spherical potential is characterized just by two quantum numbers $j$ and $m$, where $m$ is the projection of the total angular momentum $j$ for a fixed radial quantum number \cite{Landau1981}. There are two quantum numbers, because the angular motion of the particle is described by two spherical angles $\theta$ and $\varphi$. On the other hand, the rigid body rotation is represented by three Euler angles $\alpha$, $\beta$ and $\gamma$ and the additional angular degree of freedom is reflected by an extra quantum number $n$ resulting in $\ket{jmn}$. 

For fixed $j$  the $\left(2j+1\right)^2$ eigenstates $\ket{jmn}$ form a multiplet, so the generators may be expressed in the block diagonal form 
\begin{align}
\label{L-expansion}
\hat{\mathbf{L}}&=\sum_{j}\ket{jmn}\mathbf{T}^j_{m^{\prime}m}\bra{jm^{\prime}n} \, ,\\
\label{R-expansion}
\hat{\mathbf{R}}&=\sum_{j}\ket{jmn}\mathbf{T}^j_{nn^{\prime}}\bra{jmn^{\prime}} \, . 
\end{align}

In terms of state vectors $\ket{jmn}$ one can define the  body and space operators of the rigid rotor $\Theta^R_g$ and $\Theta^L_g$, respectively, satisfying
\begin{subequations}
	\begin{align}
	\Theta^R_g \ket{jmn} &=
	\ket{jmn'} D^{j}_{n'n}\left(g\right)  
	=
	\ket{jmn'} D^{j}_{n'n}\left(\alpha,\beta,\gamma\right) \, , \label{eq:UnitaryLeft} \\  
	\Theta^L_g \ket{jmn} &=
	D^{j}_{mm'}\left(g\right)\ket{jm'n} 
	=
	D^{j}_{mm'}\left(\alpha,\beta,\gamma\right)\ket{jm'n}  \,.  \label{eq:UnitaryRight}
	\end{align}
\end{subequations}

- out of which we can rename the $R_a$ and $L_a$ operators right and left generators, respectively, generating right and left group transformations.	

The state vectors $\ket{jmn}$  form a full set of eigenstates of the  total angular momentum operator \eqref{JRL}. In order to obtain their wavefunctions in coordinate space, we introduce the dual basis of 
 \emph{group elements} $\ket{g}$  or parameters $\ket{\alpha, \beta, \gamma}$. The transition between the bases gives the wavefunctions of the free isotropic rigid rotor
 \begin{equation}
 \left\langle \alpha\beta\gamma | jmn \right\rangle =
 \sqrt{\frac{2j+1}{8\pi^2}}D^{j}_{mn}\left(\alpha,\beta,\gamma\right) \,.
 \end{equation}
 as solved by Wigner \cite{Wigner1959}. 

As in the U(1) case  we introduce the \emph{group element operator} as the operator inducing a representation on the states $\ket{\alpha, \beta, \gamma}$. These operators, which will later be used as the  \emph{gauge connections},
are defined by
\begin{align}
\hat{U}^j_{mn} \ket{\alpha, \beta, \gamma} = D^{j}_{mn}(\alpha, \beta, \gamma) \ket{\alpha, \beta, \gamma} \,.
\label{Udef}
\end{align}
Here $\hat{U}^j_{mn}$ is a $\left(2j+1\right)\times\left(2j+1\right)$ matrix of operators acting on the rigid rotor Hilbert space.
From the definition of the gauge connection we obtain immediately
\begin{equation}
	[\hat{U}^j_{mn},\hat{U}^{j'}_{m'n'} ]=0
\end{equation}
and 
\begin{equation}
	\hat{U}^j_{mn} (\hat{U}^j)^{\dagger}_{nk} = \delta_{mk}\, .
\end{equation}
Using the definition of the unitary operators \eqref{eq:UnitaryLeft} and \eqref{eq:UnitaryRight} the gauge connection $\hat{U}^j_{mn}$ transforms as
\begin{subequations}
	\begin{align}
	\Theta^R\left(\alpha,\beta,\gamma\right)
	U^j_{mn}
	\Theta^{R\dagger}\left(\alpha,\beta,\gamma\right)
	&=U^j_{mn'}D^j_{n'n}\left(\alpha,\beta,\gamma\right) \,,  \\
	\Theta^L\left(\alpha,\beta,\gamma\right)
	U^j_{mn}
	\Theta^{L\dagger}\left(\alpha,\beta,\gamma\right)
	&=D^j_{mm'}\left(\alpha,\beta,\gamma\right)U^j_{m'n} \, ,
	\end{align}
\end{subequations}
{leading, by considering infinitesimal transformations, to}
\begin{subequations}
\begin{align}
[\hat{R}_a,\hat{U}^j_{mn}] &= \hat{U}^j_{mn'}(T_a^j)_{n'n} \, , \\
	[\hat{L}_a,\hat{U}^j_{mn}] &= (T_a^j)_{mm'} \hat{U}^j_{m'n} \, .
\end{align}
\end{subequations}
{The transformation laws and commutation relations for the non-Abelian case appear
as an analogy to the U(1) case. In the Abelian case the $\hat{U}$ operator raises the electric field, which is manifested by $\left[\hat{E},\hat{U}\right]=\hat{U}$. The previous equations are the non-Abelian generalization of the transformation laws and commutation relations. U(1) is an Abelian group whose irreducible representations are one dimensional, and therefore the representation states are labeled only by the quantum number $j$ (no $m$ and $n$ are needed). Raising $\ket{j}$ to  $\ket{j+1}$ by $\hat{U}$ could be seen as combining the representations $j$ and $1$ of U(1) to $j+1$.}

{The definition of the gauge connection in Eq.~\eqref{Udef} allows 
to derive the action of the operator $\hat{U}^{j}$ on the states $\ket{jmm'}$, where the states $\ket{jmm'}$ are the eigenstates defined by Eq.~\eqref{eq:DefinitionOfJKM}. The explicit expression of $\hat{U}^{j}$ is given by
\begin{align}
	\hat{U}^j_{mm'} &= \sum_{J,K} \sqrt{\frac{\dim\left(J\right)}{\dim(K)}}\left\langle JMjm | KN \right\rangle \left\langle KN' | JM'jm' \right\rangle  \ket{KNN'}\bra{JMM'} \, .
	\label{U^j_mm'}
\end{align}
A proof of this relation is given in \cite{Zohar2015}. If $\hat{U}_{mm'}^j$ acts on 
a $\ket{JMM'}$ state, the resulting states  $\ket{KNN'}$ are those reachable by 
adding angular momentum $j$ to the state $\ket{JMM'}$. The addition of angular 
momentum manifests in the Clebsch-Gordan coefficients $\braket{JMjm | KN} $ 
and $\braket{ KN' | JM'jm'} $. In particular, the 
structure of $\hat{U}^j_{mm'}$  makes sure that both the initial and final states 
of the body and space frame remain in the same representation.
}
 The state with $j=0$, $\ket{000}$, is called the singlet state, as it is invariant under all the group transformations - $\hat{\Theta}^R\left(\alpha,\beta,\gamma\right)\ket{000}=\hat{\Theta}^L\left(\alpha,\beta,\gamma\right)\ket{000}=\ket{000}$. The states $\ket{jmn}$ can be created by acting on it with the group element operator \cite{Zohar2015}
	\begin{equation}
	\ket{jmn} = \sqrt{\dim\left(j\right)} \hat{U}^j_{mn}\left|000\right\rangle \,.
	\end{equation}	

%
	
\subsection{Matter degrees of freedom: SU(2) spinor representation}
The last ingredient in order to discuss the SU(2) lattice gauge theory is the matter - fermions. Therefore we introduce fermionic creation operators $\hat{\psi}^{j\dagger}_m$, that form the elements of a \emph{spinor} in the $j$ representation. The number of components is the dimension of the representation - $2j+1$ in the case of SU(2). In the important case of the fundamental representation ($j=1/2$) such a spinor has two components, which we label with $m=1,2$. Again 
we define unitary transformations of the spinor such that
\begin{equation}
\hat{\psi}_m \rightarrow 
 D_{mm'} \left(\alpha,\beta,\gamma\right) \hat{\psi}_{m'}
\end{equation}
- this transformation mixes the elements of the spinor, just like a conventional $j$ multiplet.
The transformation of the fermions is generated by the set of generators
\begin{equation}
\hat{Q}_a = \hat{\psi}^{j\dagger}_m \left(T_a\right)_{mn} \hat{\psi}_n \, ,
\end{equation}
which fulfill the Lie algebra of the group:
\begin{equation}
\left[\hat{Q}_a,\hat{Q}_b\right]=i\epsilon_{abc} \hat{Q}_c \,.
\end{equation}
Since the operators $\hat{Q}_a$ fulfill the SU(2) algebra, one can use $\hat{\mathbf{Q}}^2=\hat{Q}_x^2+\hat{Q}_y^2+\hat{Q}_z^2$ and $\hat{Q}_z$ as 
maximal set of commuting operators to characterize fermionic states. 
For example, consider the fundamental representation $j=1/2$, where the charges are defined via the Pauli matrices
\begin{equation}
\hat{Q}_a = \frac{1}{2} \hat{\psi}^{\dagger}_m \left(\sigma_a\right)_{mn} \hat{\psi}_n \,.
\end{equation}
Note that we omitted the index $j$, which we will suppress whenever the fundamental representation is discussed.
In the following we discuss the transformation properties of the fermionic states created from the vacuum $\ket{\mathrm{vac}}_F$. The four possible states are:
\begin{itemize}
	\item The vacuum - $\ket{\mathrm{vac}}_F$ - a singlet state: 
\begin{subequations}
	\begin{align}
	&\hat{\mathbf{Q}}^2 \ket{\mathrm{vac}}_F = 0 \, , \\
	&\hat{Q}_z \ket{\mathrm{vac}}_F = 0 \, .
	\end{align}
\end{subequations}
	\item The two single fermion states - forming a fundamental, $j=1/2$ multiplet,
\begin{subequations}
	\begin{align}
	&\hat{\mathbf{Q}}^2\hat{\psi}^{\dagger}_m\ket{\mathrm{vac}}_F = \frac{1}{2}\left(\frac{1}{2}+1\right)\hat{\psi}^{\dagger}_m\ket{\mathrm{vac}}_F \, ,\\
	&\hat{Q}_z\hat{\psi}^{\dagger}_m\ket{\mathrm{vac}}_F = m \hat{\psi}^{\dagger}_m\ket{\mathrm{vac}}_F \, .
	\end{align}
\end{subequations}
	\item The doubly occupied state is a singlet:
\begin{subequations}
	\begin{align}
	&\hat{\mathbf{Q}}^2\left(\frac{1}{2}\epsilon_{mn}\hat{\psi}^{\dagger}_m\hat{\psi}^{\dagger}_n\right)\ket{\mathrm{vac}}_F = 0 \, ,\\
	&\hat{Q}_z\left(\frac{1}{2}\epsilon_{mn}\hat{\psi}^{\dagger}_m\hat{\psi}^{\dagger}_n\right)\ket{\mathrm{vac}}_F = 0 \, ,
	\end{align}	
\end{subequations}	
	where $\epsilon_{ij}$ is the $2\times 2$ antisymmetric tensor. The doubly occupied state is created by adding two fermions with $j=1/2$, each with an opposite spin which can add up to $j=0$ and $j=1$. However, the fermionic statistics only allow for the antisymmetric $j=0$ representation as can be checked by direct inspection.
\end{itemize}

\subsection{A non-Abelian one link theory: the case of SU(2)}\label{Subsec:H-SU(1)-1link} 

\begin{figure}\centering{\includegraphics[width=\columnwidth]{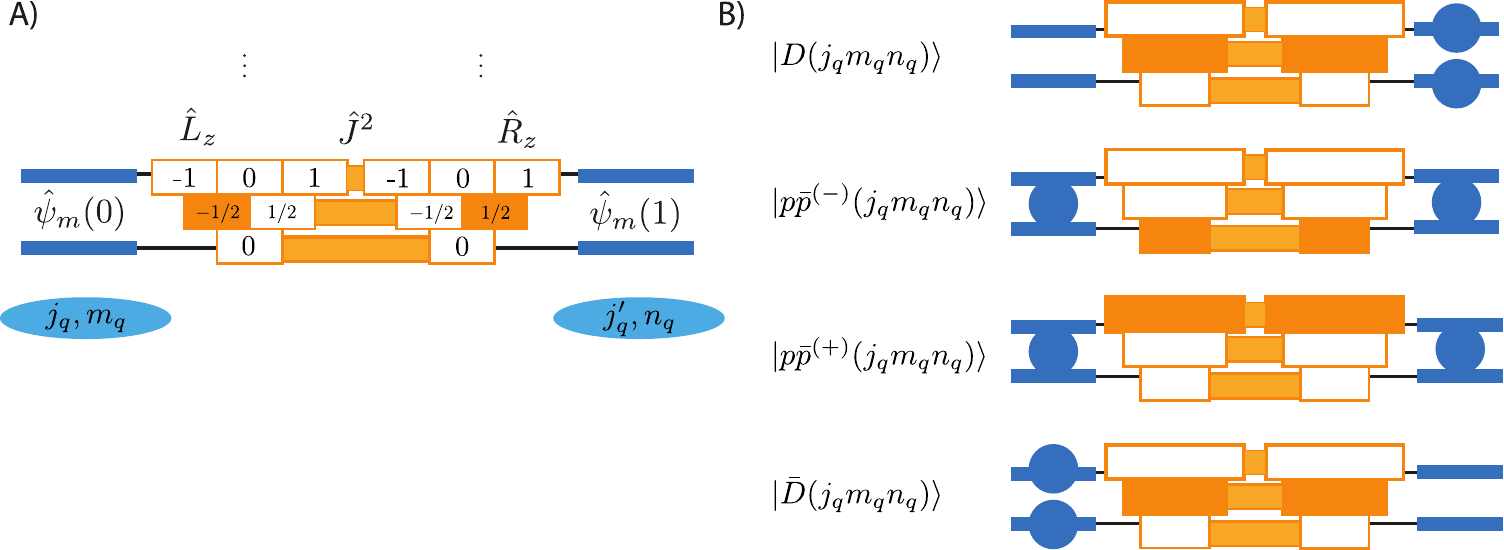}
\caption{\label{Fig:coloredJC} Visualization of the sites with non-Abelian coupling. The matter field sites consist of two fermionic components (blue lines). The link is spanned up by two spin ladders $L$ and $R$, which always have to have the same representation (spin length). B) The three states, which are part of the ground state manifold.}
}
\end{figure}

In this subsection we formulate a single link theory with a local non-Abelian gauge invariance. 
The Hamiltonian takes the form 
\begin{equation}
\label{eq:H-SU(1)-1link}
H=\frac{g^2}{2}\hat{\mathbf{J}}^2 + M [
\hat{\psi}^{\dagger}_m(0)\hat{\psi}_m(0) - \hat{\psi}^{\dagger}_m(1)\hat{\psi}_m(1)] + 
\epsilon[\hat{\psi}^{\dagger}_m(0) \hat{U}_{mn} \hat{\psi}_n(1) + h.c.] \, 
\end{equation}
with $\hat{U}_{mn}\equiv \hat{U}^{j'=1/2}_{mn}$. {A graphical illustration of the Hamiltonian Eq.~\eqref{eq:H-SU(1)-1link} can be found in Fig.~\ref{Fig:coloredJC}.} As indicated in the paragraph above Eq.~\eqref{U^j_mm'}, the SU(2) operator $\hat{U}^{j'}$ adds $j'$  to the existing representation $j$, thus changing the rotational energy of the system. Hence the operator $\hat{U}_{mn}$ featured in the Hamiltonian \eqref{eq:H-SU(1)-1link}  describes changes of the rotation quantum number $j$ by $1/2$ after the fermion undergoes a transition from the site $1$ to the site $0$. In this way $\hat{U}_{mn}$ plays a similar role as the operator $\hat{U}$ featured in the Hamiltonian \eqref{H_U1} for the  U(1) link model.

%
The Hamiltonian \eqref{eq:H-SU(1)-1link} is invariant under the local (gauge) transformations
\begin{subequations}
\begin{align}
&\hat{\psi}_m\left(0\right) \rightarrow
 D_{mm'}(g(0))\hat{\psi}_{m'}\left(0\right), \hat{U}_{mn} \rightarrow
 D_{mm'}(g(0))\hat{U}_{m'n} \, ,\\
 &\hat{\psi}_m\left(1\right) \rightarrow
 D_{mm'}(g(1))\hat{\psi}_{m'}\left(1\right), 
\hat{U}_{mn} \rightarrow
\hat{U}_{mn'}D^{\dagger}_{n'n}(g(1)) \, ,
\end{align}
which are generatated by
\end{subequations}
\begin{align}
\hat{G}_a\left(0\right)&=
\hat{L}_a-\hat{Q}_a\left(0\right) \, ,\\
\hat{G}_a\left(1\right)&=
-\hat{R}_a-\hat{Q}_a\left(1\right) \, 
\end{align}
- at each site $x$ exists a set of non-commuting generators - three for SU(2) satisfying the left algebra of the group
\begin{equation} 
\left[\hat{G}_a(x), \hat{G}_b(x)\right]=-i\epsilon_{abc} \hat{G}_c(x) \quad \forall x\,,
\end{equation}
and hence it is impossible to diagonalize all three generators with a common
basis. Moreover, the generators of the symmetry commute with the Hamiltonian 
\begin{equation}
\left[\hat{G}_a\left(x\right),\hat{H}\right] = 0 \quad \forall x,a \, .
\end{equation}

First, we consider the sector with no static charges, that is, the sector whose states $\ket{\phi ( \{0\})}$ are all strictly invariant under gauge transformations,
\begin{equation}
S_g(x)\ket{\phi ( \{0\})} = \ket{\phi ( \{0\})} \quad \forall g,x \, .
\end{equation}
In this sector, and only in this sector, all generators commute, because of all Gauss's laws
\begin{equation}
\hat{G}_a\left(x\right) \ket{\phi ( \{0\})} =0\,,
\end{equation}
which implies that the divergence of non-Abelian electric fields - left from one side and right from the other - equals exactly the dynamic charge at the vertex.

Further, we introduce static charge localized on the vertex $x$, which implies that the action of $\hat{G}_a\left(x\right)$ on such a charged state is not zero. However, it is not possible to diagonalize all the generators simultaneously, which makes it necessary to label the states by nonzero quantum numbers determined  by a maximal set of commuting operators at the vertex - $\hat{\mathbf{G}}^2\left(x\right)$ and $\hat{G}_z\left(x\right)$. This results in the usual multiplet structure: static charges form multiplets, associated with a representation $j$ of SU(2). Each site of our lattice therefore is associated with labels $j_q\left(x\right),m_q\left(x\right)$ specifying the multiplet of local static charges, and the position within the multiplet. We therefore label our static charge sectors by a collection $\left\{j_q\left(x\right),m_q\left(x\right)\right\}$, and the Gauss's laws in a given sector takes the form
\begin{equation}
\hat{G}_a\left(x\right)\left|\hat{\phi}\left(\left\{j_q\left(x\right),m_q\left(x\right)\right\}\right)\right\rangle =
(T^{j_q\left(x\right)}_a)_{m'_q\left(x\right),m_q\left(x\right)}
\ket{\psi(\{j_q\left(x\right),m'_q\left(x\right)\})}
\end{equation}
- the generators of gauge transformations do not leave the states invariant in general, but rather mix the elements of the so-called \emph{static charge multiplet}.

If a vertex contains no static charge ($j_q=0$), the action of $\hat{G}_a$ on this
vertex will result in $T_a^0=0$ indicating invariance as expected. If an additional vertex possesses a fundamental static charge $j_q=1/2$, two different $m_q$ values will be necessary. Acting with the operators $\hat{G}_a$ on such a states will rotate the states by  $T_a = \frac{1}{2}\sigma_a$. Consequently the action of $\hat{G}_z$ results in two eigenstate equations (for the two different $m_q$ states) with eigenvalues $\pm 1/2$. The $\hat{G}_x,\hat{G}_y$ equations, on the other hand, will mix these two $\hat{G}_z$ eigenstates with respect to the matrix elements of $\sigma_x/2$ and $\sigma_y/2$, respectively.

To formalize this procedure one defines a \emph{Hilbert space sector} as the set of all states which share the same $\hat{\mathbf{G}}^2\left(x\right)$ eigenvalues everywhere, and label it by the set of the represenation indices $\left\{j_q\left(x\right)\right\}$. A sector is closed under the dynamics and under all possible gauge transformations. We define a \emph{Hilbert space subsector} as the subset of states within a sector, that also share the same $\hat{G}_z\left(x\right)$ eigenvalues everywhere - labeled by $\left\{j_q\left(x\right),m_q\left(x\right)\right\}$. A subsector is closed by the dynamics, but is invariant only under gauge transformations generated by $\hat{G}_z$: arbitrary gauge transformations mix the subsectors within a sector. Moreover, the Hamiltonian or more generally every gauge invariant operator is block diagonal in the subsectors - since such operators commute with all the $\hat{\mathbf{G}}^2,\hat{G}_z$ operators. Furthermore,
the matrix elements of any gauge invariant operator will be identical for all the subsectors within a given sector 

\subsection{The single link case: no static charges}

In order to understand the spectrum of the single link Hamiltonian we exploit
the Gauss's law at the vertices i.e.
\begin{align}
\hat{L}_a\ket{\phi} & = \hat{Q}_a(0) \ket{\phi} \, , \\
\hat{R}_a\ket{\phi} &= -\hat{Q}_a(1) \ket{\phi} \, ,
\end{align}
which yield
\begin{align}
\hat{\mathbf{L}}^2 \ket{\phi} & = \hat{\mathbf{Q}}^2(0) \ket{\phi}\,, \\
\hat{\mathbf{R}}^2 \ket{\phi} & = \hat{\mathbf{Q}}^2(0) \ket{\phi}\,.
\end{align} 
Since $\hat{\mathbf{L}}^2 = \hat{\mathbf{R}}^2 = \hat{\mathbf{J}}^2$, we obtain that the representation of the link is equal to the representation on any of the vertices.

At half-filling the first option includes the Dirac sea state for the fermions - the right site $x=1$ is fully occupied, and the left site ($x=0$) is empty and the
Gauss's law implies the gauge field to be $\left|000\right\rangle$. The entire
state becomes
\begin{equation}
\ket{D} = \psi_1^{\dagger}(1)\psi_2^{\dagger}(1) \ket{\mathrm{vac}}_F \otimes \left|000\right\rangle = \frac{1}{2}\epsilon_{mn}\psi_m^{\dagger}(1)\psi_n^{\dagger}(1) \ket{\mathrm{vac}}_F \otimes \left|000\right\rangle \,.
\end{equation}

The second possibility to distribute the fermions is one fermion per site. 
In principle there are four configurations, however, gauge invariance forces to create this state from $\left|0\right\rangle$ by using a gauge invariant operator. One possibility is to use $\hat{\psi}^{\dagger}_m(0) U_{mn}\psi_n(1)$, which produces only one particular superposition of the four states, which
will be gauge invariant. Normalizing the state results in 
\begin{equation}
\ket{p\bar{p}} = \frac{1}{\sqrt{2}}\hat{\psi}^{\dagger}_m(0) U_{mn}\psi_n(1) \ket{D} =
\frac{1}{2} \hat{\psi}^{\dagger}_m(0) \epsilon_{nk} \psi^{\dagger}_k(1) \ket{\mathrm{vac}}_F \otimes \ket{\frac{1}{2}mn} \,.
\end{equation}

The third option consists of having two fermions on the left site, where both sites are in the $j=0$ representation. No particles or anti-particles are present and the link is not excited. The state is given by
\begin{equation}
\ket{\bar{D}}= \hat{\psi}_1^{\dagger}(0)\hat{\psi}_2^{\dagger}(0) \ket{\mathrm{vac}}_F \otimes \left|000\right\rangle = \frac{1}{2}\epsilon_{mn}\hat{\psi}_m^{\dagger}(0)\hat{\psi}_n^{\dagger}(0) \ket{\mathrm{vac}}_F  \otimes \left|000\right\rangle \, .
\end{equation}

The three states span the Hilbert space sector, because of the gauge symmetry. If we order the basis of our sector as $\ket{D},\ket{p\bar{p}},\ket{\bar{D}}$, the matrix representation of the Hamiltonian restricted to this sector is given by
\begin{align}
H|_{0} &= \left(\begin{array}{ccc}
-2M &\sqrt{2}\epsilon &0\\
\sqrt{2}\epsilon& \frac{3g^2}{8}&\sqrt{2}\epsilon\\
0 & \sqrt{2}\epsilon&2M
\end{array}\right) \, .
\label{H|_(0)}
\end{align}
{The spectrum and the dynamics generated by this Hamiltonian is illustrated in Fig.~\ref{Fig:coloredJCEnergies}.}

{A possible experimental realization of the Hamiltonian \eqref{H|_(0)} can be 
achieved by using alkali atoms in the $F=1$ manifold. 
In particular, one has to engineer an effective magnetic field $\mathbf{B}_{\mathrm{eff}}\propto (2\epsilon,0,-2M)$ generated by Raman or IR coupling, and the term $3g^2/8$ is described by the quadratic Zeeman effect \cite{Lin2011, Goldman2014}.}

In the case of zero coupling ($\epsilon=0$) between fermions and gauge fields, 
the basis states $\ket{D},\ket{p\bar{p}},\ket{\bar{D}}$ are the exact eigenstates, and are visualized in Fig.\,\ref{Fig:coloredJCEnergies}. $\ket{D}$, the ground state in the absence of interactions, involves the Dirac sea (no particles and anti-particles) in a product with the gauge field singlet state. The state $\ket{p\bar{p}}$ is a \emph{meson}, an excitation which includes a particle and an anti-particle connected by an electric \emph{flux tube} in a gauge invariant fashion. Finally, $\ket{\bar{D}}$ includes two particles on the left site and two anti-particles (the absence of fermions) on the right one, and may be interpreted as a pair of a \emph{baryon} and an \emph{anti-baryon} in our toy model. A baryon is an antisymmetrized (and therefore colorless) combination of particles of all possible colors, (the particles on the left vertex), and the right site realizes the anti-baryon. Since baryons are color neutral, they are singlets and need not be connected by any flux tubes, manifested by the absence of gauge field excitations in $\ket{\bar{D}}$. With the increase of the interaction strength $\epsilon$, the vacuum becomes more and more dressed by the \emph{meson} state $\left|p\bar{p}\right\rangle$. At $M \approx \frac{3g^2}{16}$ the meson state $\left|p\bar{p}\right\rangle$ becomes resonant with the baryon-antibaryon state $\ket{\bar{D}}$. The resulting resonant dynamics are sketched in Fig.~\ref{Fig:coloredJCEnergies}~B.

\subsection{Charging up the link: static charges}

\begin{figure}\centering{\includegraphics[width=\columnwidth]{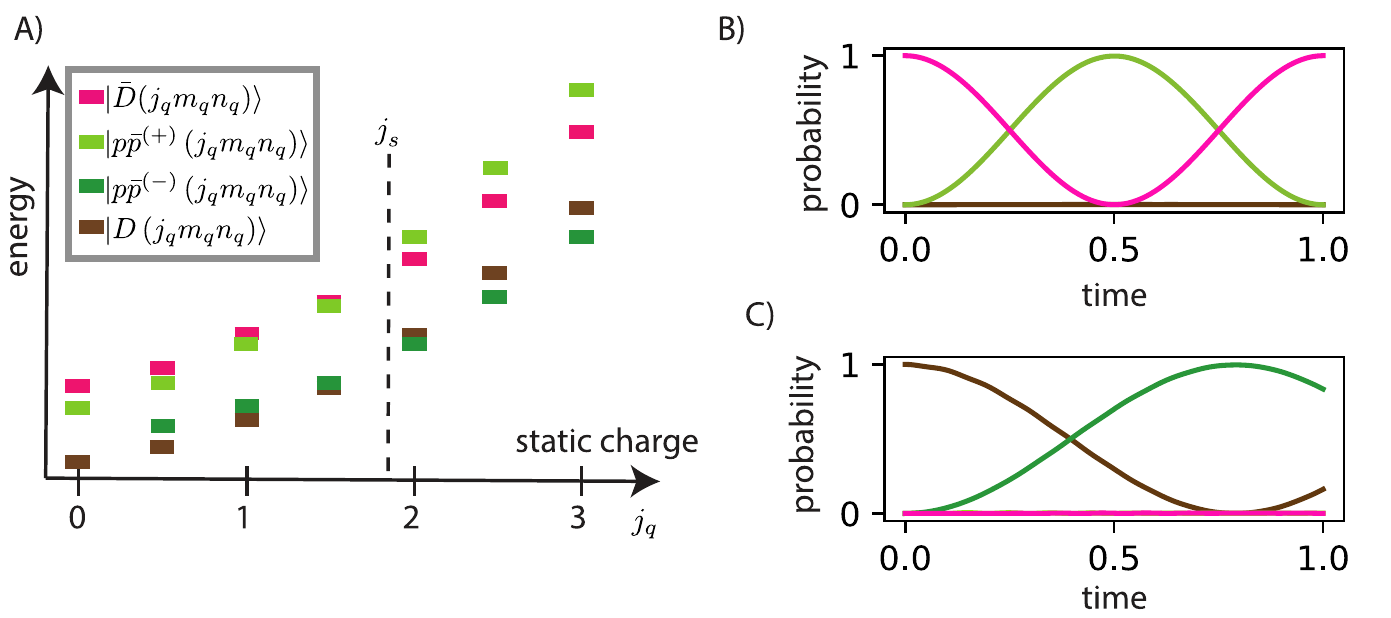}
\caption{\label{Fig:coloredJCEnergies} Analysis of SU(2) with a single link. A) The spectrum for $g= 1.5$ and $M=2$. B) Analog of quantum Rabi oscillations for SU(2).  The $\ket{\bar{D}\,(000)}$ couples resonantly into the state $\ket{p\bar{p}\,(000)}$, while the state  $\ket{D\,(000)}$ remains empty for $M=2\pi\cdot3$,
$g = \sqrt{2\pi}4$  and $\epsilon = \frac{2\pi}{\sqrt{8}}$. C) Instability of the $\ket{D\,(j_qm_qn_q)}$ state induced by the existence of static charges. The state $\ket{D\,(2m_qn_q)}$ couples resonantly to $\ket{p\bar{p}^{(-)}\,(2m_qn_q)}$, while the other two states remain mostly empty for $M=2\pi\cdot 2.25$,
$g = \sqrt{2\pi}2$  and $\epsilon = \frac{2\pi}{\sqrt{8}}$.}
}
\end{figure}

In the next step we introduce static charges into the single link model. Suppose we place on the left site some static charge with $j_q,m_q$ and $j_q>0$, and another static charge $j'_q,n_q$ on the right site, which also determine the subsector.
As before, the number of fermionic possibilities is highly restricted. We begin with the configuration that includes the Dirac sea - both fermions are on the right site denoted by $\ket{0(j_q,m_q,j'_q,n_q)}$ a state which satisfies that. The fermion configuration determines the eigenvalues of the charge operators
\begin{equation}
\hat{Q}_a(0)\ket{0(j_q,m_q,j'_q,n_q)}
=
\hat{Q}_a(1)\ket{0(j_q,m_q,j'_q,n_q)} = 0 \, 
\end{equation}
and the Gauss's law implies that
\begin{align}
\hat{L}_a\ket{0(j_q,m_q,j'_q,n_q)} &= \hat{G}_a(0)\ket{0(j_q,m_q,j'_q,n_q)} \, ,\\
-\hat{R}_a\ket{0(j_q,m_q,j'_q,n_q)} &= \hat{G}_a(1)\ket{0(j_q,m_q,j'_q,n_q)} \, .
\end{align}
Due to $\hat{\mathbf{L}}^2(0)=\hat{\mathbf{R}}^2(0)$ we conclude that 
\begin{equation}
\hat{\mathbf{G}}^2(0)\ket{0(j_q,m_q,j'_q,n_q)} = \hat{\mathbf{G}}^2(1)\ket{0(j_q,m_q,j'_q,n_q)}
\end{equation}
and since the eigenvalues of all $\hat{\mathbf{G}}^2$ are fixed within the sector, a similar condition applies to all the states of the sector	and we conclude that $j_q = j'_q$ in the one link case.
The Gauss's law also dictates exactly the state of the gauge field for $\left|0\left(j_q,m_q,n_q\right)\right\rangle$, and we conclude that there is only one such state per subsector
\begin{equation}
\ket{D\left(j_q m_q n_q\right)} =
\frac{1}{2}\epsilon_{mn}\hat{\psi}^{\dagger}_m(1) \hat{\psi}^{\dagger}_n(1) \left|0\right\rangle_F \otimes \ket{j_q, m_q, -n_q} 
\end{equation}
- a state with Dirac sea and an electric field originating from the static charges on top of it.

Employing again the Gauss's law we can define a state for the case where both fermions are on the right site
\begin{equation}
\ket{\bar{D}\left(j_q m_q n_q\right)} =
\frac{1}{2}\epsilon_{mn}\hat{\psi}^{\dagger}_m(0)\hat{\psi}^{\dagger}_n(0) \left|0\right\rangle_F \otimes \ket{j_q, m_q, -n_q}
\end{equation}
- representing a baryon-antibaryon pair, again, but this time with the electric field originating from the static charges.

Finally, we have to consider the action of the mesonic operator on any of these two states, to obtain the other basis elements of the subsector. Acting on $\ket{0\left(j_q m_q n_q\right)}$, we obtain
\begin{equation}
\hat{\psi}^{\dagger}_m(0) \hat{U}_{mn}(0) \hat{\psi}_n(1)\left|0\left(j_q m_q n_q\right)\right\rangle =
 \hat{\psi}^{\dagger}_m(0) \epsilon_{nk} \hat{\psi}^{\dagger}_k(1) \ket{\mathrm{vac}}_F \otimes \hat{U}_{mn}\ket{j_q, m_q, -n_q} \,.
\end{equation}
This creates a superposition of two states with two different gauge field representations for $j_q \pm \frac{1}{2}$ resulting in
\begin{align}
\hat{U}^{1/2}_{mn}(0) \ket{j_q, m_q ,-n_q} &=
 C^{j_q+\frac{1}{2}}_{mn} \ket{j_q +\tfrac{1}{2}, m_q+m, -n_q+n} + C^{j_q-\frac{1}{2}}_{mn} \ket{j_q -\tfrac{1}{2}, m_q+m, -n_q+n}
\end{align}
where
\begin{subequations}
\begin{align}
C^{j_q+\frac{1}{2}}_{mn} &= \sqrt{\frac{2j_q+1}{2j_q+2}} \braket{j_q m_q; \frac{1}{2} m|j_q + \frac{1}{2}, m_q+m} \braket{j_q+\frac{1}{2}, -n_q+n|j_q, -n_q; \frac{1}{2} n} \, ,\\
C^{j_q-\frac{1}{2}}_{mn} &=\sqrt{\frac{2j_q+1}{2j_q}} \braket{j_q m_q; \frac{1}{2} m|j_q - \frac{1}{2}, m_q+m} \braket{j_q-\frac{1}{2}, -n_q+n|j_q ,-n_q ;\frac{1}{2} n} \,.
\end{align}
\end{subequations}
With the proper normalizations and Clebsch-Gordan calculus we obtain that
\begin{equation}
\hat{\psi}^{\dagger}_m(0) \hat{U}_{mn}(0) \hat{\psi}_n(1)\left|0\left(j_q m_q n_q\right)\right\rangle =
\sqrt{\frac{2j_q+2}{2j_q+1}}\ket{p\bar{p}^{(+)}\left(j_q m_q n_q\right)} + \sqrt{\frac{2j_q}{2j_q+1}}\ket{p\bar{p}^{(-)}\left(j_q m_q n_q\right)} \, ,
\end{equation}
where the two remaining basis states are
\begin{equation}
\ket{p\bar{p}^{\left(\pm\right)}\left(j_q m_q n_q\right)}
	=  \hat{\psi}^{\dagger}_m(0)C^{j_q+\frac{1}{2}}_{mn}\epsilon_{nk}\hat{\psi}_k^{\dagger}(1)
	 \ket{\mathrm{vac}}_F\otimes\ket{j_q\pm\frac{1}{2} ,m_q+m,-n_q+n} \,.
\end{equation}
The restriction of the Hamiltonian to any other subsector of the $j_q>0$ sector, leads to 
\begin{equation}\label{Eq:HamSimpleSu2}
\begin{aligned}
&H|_{j_q m_q n_q} =\\& \left(\begin{array}{cccc}
-2M + \frac{g^2}{2}j_q\left(j_q+1\right)& \sqrt{\frac{2j_q+2}{2j_q+1}}\epsilon&
\sqrt{\frac{2j_q}{2j_q+1}}\epsilon &
0\\
\sqrt{\frac{2j_q+2}{2j_q+1}}\epsilon& \frac{g^2}{2}\left(j_q+\frac{1}{2}\right)\left(j_q+\frac{3}{2}\right)&
0&
\sqrt{\frac{2j_q+2}{2j_q+1}}\epsilon\\
\sqrt{\frac{2j_q}{2j_q+1}}\epsilon &
 0&
\frac{g^2}{2}\left(j_q-\frac{1}{2}\right)\left(j_q+\frac{1}{2}\right)&
 \sqrt{\frac{2j_q}{2j_q+1}}\epsilon\\
0& 
\sqrt{\frac{2j_q+2}{2j_q+1}}\epsilon&
\sqrt{\frac{2j_q}{2j_q+1}}\epsilon &
2M+ \frac{g^2}{2}j_q\left(j_q+1\right)
\end{array}\right)
\end{aligned} \, ,
\end{equation}
where the basis states are ordered as $\ket{D\left(j_q m_q n_q\right)}$, $\ket{p\bar{p}^{(+)}\left(j_q m_q n_q\right)}$, $\ket{p\bar{p}^{(-)}\left(j_q m_q n_q\right)}$, $\ket{\bar{D}\left(j_q m_q n_q\right)}$. While this matrix representation holds for $j_q>0$, one observes a decoupling of the third row and third column for $j_q=0$, and the 
the problem becomes effectively three dimensional i.e. $H|_0$ . Altogether, the Hamiltonian of the single link case is given by the direct sum of blocks:
\begin{equation}
H = \underset{j_q,m_q,n_q}{\bigoplus}H|_{j_q m_q n_q} \, ,
\end{equation} 
where $H|_{j_q m_q n_q} = H|_{j_q m'_q n'_q}$ is the same for all the subsectors belonging to the same sector $j_q $ {and the spectrum and dynamics of a single block is illustrated in Fig.~\ref{Fig:coloredJCEnergies}.}
 The constituting $4\times4$ matrix-Hamiltonians $H|_{j_q m_q n_q}$ given by Eq. \eqref{Eq:HamSimpleSu2} describe systems of two interacting qubits. Ignoring a common energy offset, we can explicitly write out the Hamiltonian \eqref{Eq:HamSimpleSu2} in terms of two interacting qubits as 
\begin{align}
\hat{H}_j &= \left(2M - \frac{g^2}{4}(2j+1)\right)\hat{s}_{z,1}
+\left( 2M + \frac{g^2}{4}(2j+1)\right)\hat{s}_{z,2}+ \sqrt{\frac{2j_q+2}{2j_q+1}}\epsilon\hat{s}_{x,1} \, \notag \\
 &+\sqrt{\frac{2j_q}{2j_q+1}}\epsilon\hat{s}_{x,2} -\left( \frac{g^2}{2}j(j+1)+\frac{g^2}{4}\right)\hat{s}_{z,1}\hat{s}_{z,2} \, ,
 \label{eq:RydbergHamiltonian}
\end{align}
where we directly see that $J < 0 $ for all $j$, hence favoring a ferromagnetic configuration.
Such a system of two interacting qubits can be implemented in quantum computing and simulation architectures like Rydberg systems, trapped ions or superconducting qubits with local control of detuning and coupling.  
{To be explicit we briefly discuss the experimental realization of the Hamiltonian Eq.~\eqref{eq:RydbergHamiltonian} with two Rydberg atoms trapped separately in an optical tweezer. Each Rydberg atom realizes a two-level system which corresponds to a spin-1/2 system and the detuning of the dressing laser from resonance allows to tune the local $\hat{s}_{z,i}$ terms. 
The locally adjustable light-dipole Rabi coupling is described by the spin-flip operator $\hat{s}_{x,i}$. Moreover, the
dipole-dipole interaction of two Rydberg atoms leads to an interaction of the form
$~\hat{s}_{z,1}\hat{s}_{z,2}$ which can be tuned by the distance of the Rydberg atoms \cite{Browaeys2020}.}
For comparison, the Abelian U(1) link provides the $2\times2$  $q$-dependent matrix Hamiltonian \eqref{eq:eigen-value-equation-U(1)link} describing a single qubit or equivalently a single spin. Thus the non-Abelian link model yields an additional complexity to quantum simulations.

In a similiar fashion as for the U(1) case in Section~\ref{Compact U(1)}, we can observe that the energies of $\ket{D,j}$ and $\ket{p\bar{p}^{(-)},j}$ invert for sufficiently strong static charges:
\begin{align}\label{Eq:SchwingerSU2}
j_s &= \frac{8M}{3g^2}-\frac{1}{2}\,.
\end{align}
This equation is the non-Abelian analog of Eq. \eqref{Eq:SchwingerPairSimple}, at which the ground state changes. We visualized this behavior in Fig.~\ref{Fig:coloredJCEnergies}~C, where the two states become energetically degenerate. In the next subsection we extend the discussion of the single link to 
a one-dimensional system of non-Abelian links.

\subsection{A $(1+1)$ dimensional SU(2) lattice gauge theory}

\begin{figure}\centering{\includegraphics[width=\columnwidth]{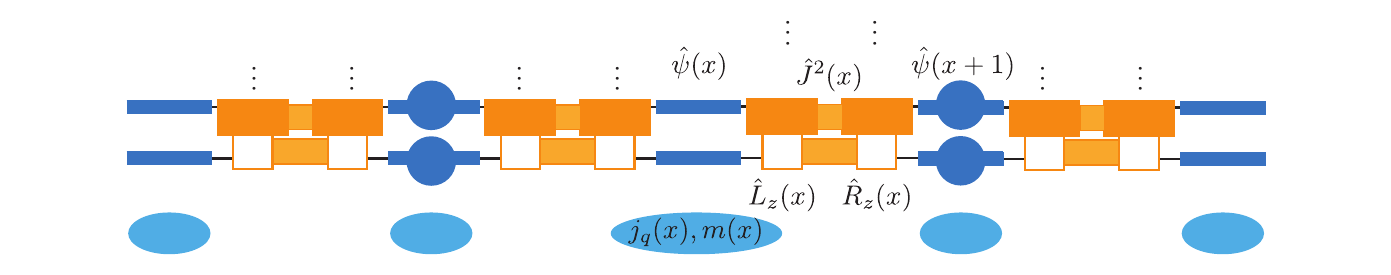}
\caption{\label{Fig:su2latticecoloredJC} The lattice formulation of SU(2). The matter field sites consist of two fermionic components (blue lines). The link is spanned up by two spin ladders $\hat{L}_a(x)$ and $\hat{R}_a(x)$, which always have to have the same representation (spin length).}
}
\end{figure}

We introduce once again a one dimensional lattice with open boundaries, with an even number of sites, $\mathcal{N}=2\mathcal{M}+2$, labeled from $0$ to $\mathcal{N}-1=2\mathcal{M}+1$, and $\mathcal{N}-1=2\mathcal{M}+1$ intermediate links. Each vertex may host at most two fermionic species, created by the operators $\{\hat{\psi}^{\dagger}_m\left(x\right)\}_{m=1,2}$, which form a fundamental SU(2) spinor. Each link hosts a rigid rotor Hilbert space. 
The Hamiltonian, which can be seen as a direct generalization of the U(1) case, see Eq. \eqref{eq:c-QED}, takes the form 
\begin{equation}
\label{H_SU2_lattice}
\hat{H}=\frac{g^2}{2}\underset{x}{\sum}\hat{\mathbf{J}}^2\left(x\right)
+M\underset{x}{\sum}\left(-1\right)^x \hat{\psi}^{\dagger}_m\left(x\right)\hat{\psi}_m\left(x\right)
+\epsilon\underset{x}{\sum} [\hat{\psi}^{\dagger}_m \hat{U}_{mn}\left(x\right) \hat{\psi}_n\left(x+1\right) + h.c. ] \, ,
\end{equation}
where we also sum over gauge-indices. In the second term the summation over $x$ covers all the lattice sites (all the vertices): $x=0,...,\mathcal{N}-1$.  On the other hand, the summation is over all the links ($x=0,...,\mathcal{N}-2$) in the first and the third terms.  In the single link case ($\mathcal{N}=2$) Eq. \eqref{H_SU2_lattice} reduces to the Hamiltonian given by Eq. \eqref{eq:H-SU(1)-1link} subject to replacement $\hat{\mathbf{J}}^2\left(0\right) \rightarrow \hat{\mathbf{J}}^2$ and $ \hat{U}_{mn} \left(0\right) \rightarrow \hat{U}_{mn}$.

The first term in the Hamiltonian~\eqref{H_SU2_lattice} corresponds to the electric energy. The second is the (staggered) mass term as before: for SU(2), staggering is simpler and is manifested only in the mass term, while the charges are defined exactly the same at each vertex with operators $\hat{Q}\left(x\right)=\psi^{\dagger}_m\left(T_a\right)_{mn}\psi_n\left(x\right)$ as defined above. The last term is the interaction between the matter and the gauge field. As in the previous section we suppress representation index $j$ when referring to the fundamental representation i.e. for SU(2) one has
$\hat{\psi}^{\dagger}_m\equiv\hat{\psi}^{1/2\dagger}_m$,
$\hat{U}_{mn}\equiv \hat{U}^{1/2}_{mn}$ and
$D_{mn}\equiv D^{1/2}_{mn}$.

A gauge transformation with respect to a set of locally chosen group element $g\left(x\right) \in SU(2)$, parametrized by Euler angles $\left\{\alpha\left(x\right),\beta\left(x\right),\gamma\left(x\right)\right\}$ acts on the various degrees of freedom as follows:
\begin{subequations}
\begin{align}
\hat{\psi}_m\left(x\right) &\rightarrow
 D_{mm'}\left(g\left(x\right)\right)\hat{\psi}_{m'}\left(x\right) \, , \\
\hat{U}_{mn}\left(x\right) &\rightarrow
 D_{mm'}\left(g\left(x\right)\right) \hat{U}_{m'n'}\left(x\right)\, ,  \\
 \hat{U}_{mn}\left(x-1\right) &\rightarrow
\hat{U}_{m'n'}\left(x\right)D^{\dagger}_{n'n}\left(g\left(x\right)\right) \,.
\end{align}
\end{subequations}
Due to the unitarity of the Wigner matrices this local transformation leaves the Hamiltonian $\hat{H}$ invariant. The gauge transformation is generated by the operators
\begin{equation}
\hat{G}_a\left(x\right) = \hat{L}_a\left(x\right) - \hat{R}_a\left(x-1\right) - \hat{Q}_a\left(x\right) 
\end{equation}
for $0<x<\mathcal{N}-1$, and
\begin{align}
\hat{G}_a\left(0\right)&=
\hat{L}_a\left(0\right)-\hat{Q}_a\left(0\right) \, ,\\
\hat{G}_a\left(\mathcal{N}-1\right)&=
-\hat{R}_a\left(\mathcal{N}-2\right)-\hat{Q}_a\left(\mathcal{N}-1\right)
\end{align}
- at each site $x$ we have a set of non-commuting generators - three in our SU(2) case, satisfying the group's left algebra commute with the Hamiltonian. 
\begin{equation}
\left[\hat{G}_a\left(x\right),\hat{H}\right] = 0 \quad \forall x,a \, .
\end{equation}

\subsection{The Hilbert Space of the One Dimensional System}
After analyzing the Hilbert space of the single link case we determine an upper bound on the dimension of the SU(2) one dimensional open system for a given static charge subsector
 $\left\{j_q\left(x\right),n_q\left(x\right)\right\}$. The system has a 
 $\mathcal{N}=2\mathcal{M}+2$ vertices labeled from $0$ to $2\mathcal{M}+1$, and the central link is located between the vertices $\mathcal{M}$ and $\mathcal{M}+1$.

The Gauss's law on the first vertex implies that the possible representations are 
given by $j_q\left(0\right),j_q\left(0\right) \pm \frac{1}{2}$ - adding the representation of the static charge on the first site with the possible representations of the dynamical matter there - $0,\frac{1}{2}$.
At the next site, we have a static charge of $j_q\left(1\right)$, implying that the next link can host all the representations from
$|j_q\left(1\right)-j_q\left(0\right)|-\frac{1}{2}$ to 
$j_q\left(1\right)+j_q\left(0\right)+\frac{1}{2}$ with half integer increments. We can move on and also perform a similar procedure from the end to the right, until we obtain that the middle link, $\mathcal{M}$, can host the maximal number of representations.

In the following we focus on states without static charges i.e. $j_q (x) = 0$ everywhere. In the case of no charges the Gauss's  law at vertex $0$ implies that the first link may host the representations $0$ and $\frac{1}{2}$.  The
same argument can be applied starting from the Gauss's law at vertex $\mathcal{N}$. Using this  recursive argument the middle link $\mathcal{M}$ can host the maximal number of representations. 
Each link $n$, there will be a finite number of allowed representations, ranging from $0$ to $J_{\text{max}}\left(n\right)$, with increments of $\frac{1}{2}$. Using both directions, we obtain that
\begin{equation}
J_{\text{max}}\left(n\right) = J_{\text{max}}\left(2\mathcal{M}-n\right)
\end{equation}
and for $0 \leq n \leq \mathcal{M}$, 
\begin{equation}
J_{\text{max}}\left(n\right) = \frac{n+1}{2} \,.
\end{equation}
Thus, we  obtain for $0 \leq n \leq \mathcal{M}$ that
\begin{equation}
\dim\left(n\right) = \underset{j=0}{\overset{J_{\text{max}}\left(n\right)}{\sum}}
\left(2j+1\right)^2 = \frac{1}{6}\left(n+2\right)\left(n+3\right)\left(2n+5\right) \, ,
\end{equation}
where the summation is over integers and half-integers. The maximal Hilbert space dimension of all the links is
\begin{equation}
\begin{aligned}
\mathcal{D}_{\text{Gauge}} \left(\mathcal{M}\right)&\equiv  \left(\prod_{n=0}^{\mathcal{M}-1}
\left(\frac{1}{6}\left(n+2\right)\left(n+3\right)\left(2n+5\right)\right)
\right)^2
\left(\frac{1}{6}\left(\mathcal{M}+2\right)\left(\mathcal{M}+3\right)\left(2\mathcal{M}+5\right)\right)
\\
&=\frac{\left(\mathcal{M}+2\right)\left(2\mathcal{M}+5\right)\left(\mathcal{M}+2\right)!\left(\mathcal{M}+3\right)!\left(2\mathcal{M}+3\right)!^2}
{2^{4\mathcal{M}+5}3^{2\mathcal{M}+3}} \, ,
\end{aligned}
\end{equation}
which is finite for a finite system size. The fermionic Fock space dimension reduces in the half filling case to $\binom{4\mathcal{M}+4}{2\mathcal{M}+2}$ - number of options to distribute $2\mathcal{M}+2$ fermions in $4\mathcal{M}+4$ positions. Therefore, we conclude that the following upper bound for the dimension
of the Hilbert space of a $(1+1)$ dimensional SU(2) lattice gauge theory   
\begin{equation}
\begin{aligned}
\mathcal{D}\left(\mathcal{M}\right)&< 
\frac{\left(\mathcal{M}+2\right)
\left(\mathcal{M}+2\right)!
\left(\mathcal{M}+3\right)!
\left(2\mathcal{M}+3\right)^2
\left(2\mathcal{M}+5\right)
\left(4\mathcal{M}+4\right)!}
{3^{2\mathcal{M}+3}4^{4\mathcal{M}+5}}
\\
&=
\frac{\left(\mathcal{N}/2+1\right)
\left(\mathcal{N}/2+1\right)!
\left(\mathcal{N}/2+2\right)!
\left(\mathcal{N}+1\right)^2
\left(\mathcal{N}+3\right)
\left(2\mathcal{N}\right)!}
{3^{\mathcal{N}+1}4^{2\mathcal{N}+1}}
\end{aligned}
\end{equation} 

This result is not independent of the charge sector, and is valid only for the sector without static charges. As in the U(1) case, it is an inequality and not an equation, since gauge invariance has not been fully exploited yet, and doing it would introduce nonlocality. Inserting $\mathcal{N}=2$ or $\mathcal{M}=0$ into the above formula for the single link case, we obtain $5$, while we know that the dimension of the $j_q=0$ sector there is $3$, from taking full gauge invariance into account. An analysis taking into account full gauge invariance is discussed in
\cite{Banuls2017} which is based on ideas from \cite{Hamer1977}

\section{Conclusion}
In this article we started from the JCM and used it to introduce a simple instance of a non-Abelian lattice gauge theory. We discussed in detail the physical Hilbert space and gave bounds for the dimension of the physical Hilbert space of a (1+1) dimensional U(1) and SU(2) lattice gauge theory with open boundary conditions. By choosing proper basis states we illustrated that the quantum simulation of the one-link non-Abelian lattice gauge theory can be achieved with a three or four level system, respectively. 

The discussion of the minimal model demonstrates two things explicitly. Firstly, the Gauss's law reduces the dimension of the Hilbert space and secondly the basis used to describe the quantum many-body Hamiltonian determines the possibilities and efficiency of quantum simulation. {Given our estimates of the main texts we expect that a 50 qubit quantum computer will be able to quantum simulate a non-Abelian system of about 12 sites and an Abelian system of about 16 sites. Hence, we consider it a particular promising next step to optimize the choice of basis state in order to facilitate the quantum simulation of non-Abelian lattice gauge theories. Similar to the $H_2$ molecule in quantum chemistry,
the non-Abelian single link is a model, which could be analyzed with current quantum computers e.g. via variational algorithms \cite{OMalley2016, Kandala2017}. These variational algorithms have shown great promise in recent studies of U(1) gauge theories \cite{Kokail2019}, and appears as a natural route towards the implementation of non-Abelian lattice gauge theories.

\section{Acknowledgments}
V.K. wants to thank Philipp Hauke and Daniel González-Cuadra for discussions on 
related work.
This work is part of and supported by the DFG Collaborative Research Centre “SFB 1225 (ISOQUANT)”. F. J. acknowledges the DFG support through the Emmy-Noether grant (project 377616843) and from the Juniorprofessorenprogramm Baden-W\"urttemberg (MWK). M.L. acknowledges the Spanish Ministry MINECO (National Plan
15 Grant: FISICATEAMO No. FIS2016-79508-P, SEVERO OCHOA No. SEV-2015-0522, FPI), European Social Fund, Fundacio Cellex, Fundacio Mir-Puig, Generalitat de Catalunya (AGAUR Grant No. 2017 SGR 1341, CERCA/Program), ERC AdG NOQIA, EU FEDER, European Union Regional Development Fund - ERDF Operational Program of Catalonia 2014-2020 (Operation Code: IU16-011424), MINECO-EU QUANTERA MAQS (funded by The State Research Agency (AEI) PCI2019-111828-2 / 10.13039/501100011033) , and the National Science Centre, Poland-Symfonia Grant No. 2016/20/W/ST4/00314. 
V.K. work was supported by the European Union under Horizon2020 (PROBIST 754510) 

\bibliography{Dyn-gauge-fields-cavity}

\end{document}